\documentclass[preprint,5p,times,authoryear,twocolumn]{elsarticle}
\usepackage{comment}
\usepackage{amsmath,amssymb,amsfonts,mathrsfs}
\usepackage{algorithmic}
\usepackage{graphicx}
\usepackage{textcomp}
\usepackage{rotating}
\usepackage{multirow}
\usepackage{lscape}
\usepackage{longtable}
\usepackage{amssymb}
\usepackage{gensymb}
\usepackage{subcaption}

\usepackage{enumitem}
\usepackage{layouts}
\usepackage{pifont}
\newcommand{\xmark}{\ding{55}}
\usepackage{tikz}
\usepackage{tabularray}
\UseTblrLibrary{booktabs}
\usepackage{array}
\usepackage{balance}
\usepackage{multicol}
\usepackage{caption}
\usepackage{url}

\usepackage{breakurl}
\usepackage[breaklinks]{hyperref}
\usepackage{jabbrv}

\makeatletter
\newcommand{\showfontsize}{The current font size is \f@size pt.}
\makeatother

\usepackage{amssymb}
\usepackage{amsmath}

\journal{AI Open}

\begin{document}

\begin{frontmatter}



\title{Advancing Histopathology with Deep Learning Under Data Scarcity: A Decade in Review}


\author[label1]{Ahmad Obeid\corref{cor1}}
\ead{ahmad.obeid@ku.ac.ae}
\cortext[cor1]{Corresponding Author}
\author[label1]{Said Boumaraf}
\ead{said.boumaraf@ku.ac.ae}
\author[label1]{Anabia Sohail}
\ead{anabia.sohail@ku.ac.ae}
\author[label2]{Taimur Hassan}
\ead{taimur.hassan@adu.ac.ae}
\author[label1]{Sajid Javed}
\ead{sajid.javed@ku.ac.ae}
\author[label1]{Jorge Dias}
\ead{jorge.dias@ku.ac.ae}
\author[label3]{Mohammed Bennamoun}
\ead{mohammed.bennamoun@uwa.edu.au}
\author[label1]{Naoufel Werghi}
\ead{naoufel.werghi@ku.ac.ae}

\affiliation[label1]{organization={Department of Electrical Engineering and Computer Science - Khalifa University},
            city={Abu Dhabi},            
            country={UAE}}

\affiliation[label2]{organization={Department of Electrical and Computer Engineering - Abu Dhabi University},
            city={Abu Dhabi},            
            country={UAE}}

\affiliation[label3]{organization={ Department of Computer Science and Software Engineering - University of Western Australia},
            city={Perth},
            country={Australia}}

\begin{abstract}
Recent years witnessed remarkable progress in computational histopathology, largely fueled by deep learning. This brought the clinical adoption of deep learning-based tools within reach, promising significant benefits to healthcare, offering a valuable second opinion on diagnoses, streamlining complex tasks, and mitigating the risks of inconsistency and bias in clinical decisions. However, a well-known challenge is that deep learning models may contain up to billions of parameters; supervising their training effectively would require vast labeled datasets to achieve reliable generalization and noise resilience. In medical imaging, particularly histopathology, amassing such extensive labeled data collections places additional demands on clinicians and incurs higher costs, which hinders the art’s progress. Addressing this challenge, researchers devised various strategies for leveraging deep learning with limited data and annotation availability. In this paper, we present a comprehensive review of deep learning applications in histopathology, with a focus on the challenges posed by data scarcity over the past decade. We systematically categorize and compare various approaches, evaluate their distinct contributions using benchmarking tables, and highlight their respective advantages and limitations. Additionally, we address gaps in existing reviews and identify underexplored research opportunities, underscoring the potential for future advancements in this field.
\end{abstract}




\end{frontmatter}



\section{Introduction}
\label{sec:introduction}
Cancer-related mortality and morbidity have risen due to aging and population growth \citep{morbid}. Despite better diagnostic and treatment technologies improving survival rates, the cancer burden remains substantial. The World Health Organization's 2016 report highlights cancer as the leading contributor to global disease burden, accounting for 244.6 million Disability-Adjusted Life Years (DALYs), and as the second highest cause of death worldwide, after ischemic heart disease, resulting in 8.97 million fatalities~\citep{who}. In the United States, however, cancer mortality rates have decreased by 25\% from 1990 to 2015, particularly in breast and colorectal cancers, largely thanks to new screening and intervention methods~\citep{intervention}. 


\subsection{Digital and Computational Pathology}
\label{section:intro-digital}

The introduction of whole-slide imaging (WSI) has enabled Digital Pathology by digitizing glass slides, facilitating a wide range of clinical, educational, and research applications. Advances in optical imaging allow for the rapid creation of extensive microscopic data~\citep{wsi}, known as virtual microscopy. Additionally, improved storage technologies support managing large datasets. Consequently, WSI devices have become indispensable in pathology for both diagnostic and research purposes. Currently, manual diagnosis remains the gold standard in pathology. However, Computational Histopathology (CPath) is rapidly advancing within Digital Pathology, aided by the integration of the ever-developing machine learning (ML) and deep learning (DL) tools. These technologies have demonstrated the capability to match or even surpass human performance in tasks such as image classification, as well as object detection and segmentation in natural scenes.


\begin{figure*}[t]
    \centering
    \includegraphics[width=\linewidth]{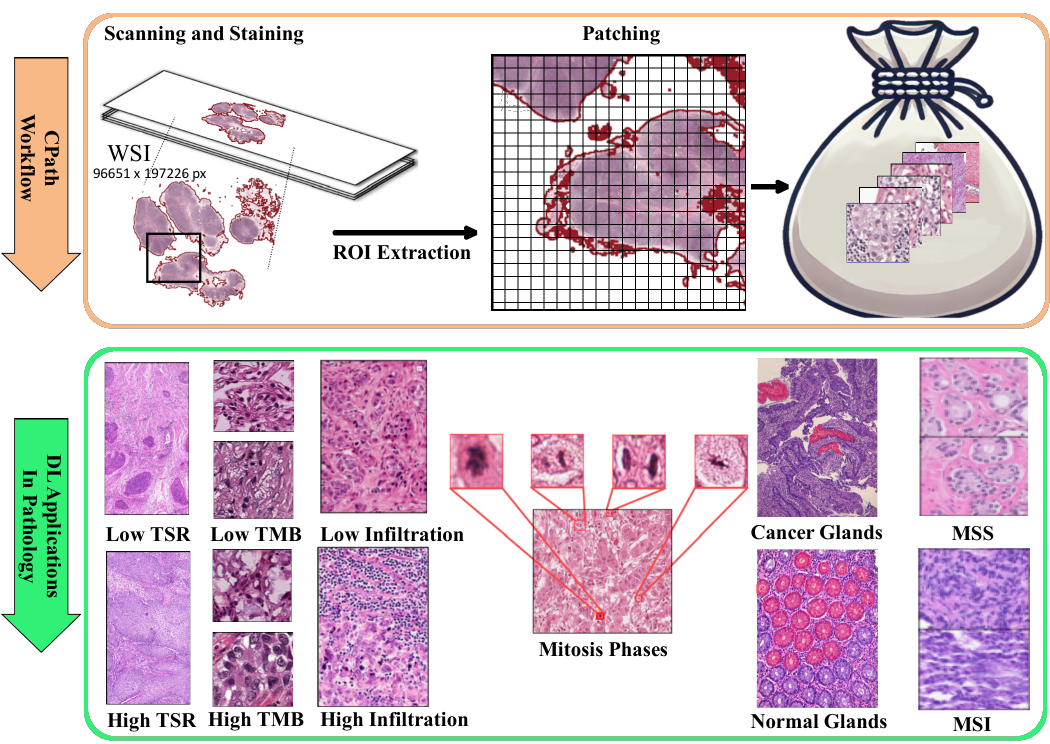}
    \caption{A pictorial illustrations of CPath. Upper: the typical CPath workflow, involving scanning, staining, ROI extraction, patch extraction, and collecting a bag of multiple instances. Lower: different cancer biomarkers where DL can be used. From left to right:  Stroma-poor vs stroma-rich lung squamous cell carcinoma groups; High-TMB vs low-TMB regions; Low vs high tumor-infiltration; Mitotic cells at different mitotic phases; Normal vs cancerous glandular structures highlighted; Microsattelite stable vs high instability (adapted from \citep{MSI3}).}
    \label{fig:apps}
\end{figure*}

Successfully applying DL capabilities to WSI analysis could substantially aid clinicians and patients by offering a reliable second opinion or assisting tool, thus addressing issues like fatigue, and heavy workloads, especially as pathologists experience rising consultation loads and a shrinking workforce~\citep{manualCRCbad}. DL tools can also reduce risks of inter-observer variability and bias, enhance diagnostic accuracy, and lead to innovative treatment approaches. 

DL tools are now transforming many tasks traditionally performed manually by pathologists. These include classifying patient biopsies, categorizing pathological WSIs, segmenting glands and tumor stroma, delineating tumor cell membranes, identifying and classifying nuclei, and detecting mitotic figures~\citep{92,breastcls,breastseg,crctasks,glas}. DL also supports downstream clinical applications such as patient survival forecasts, cancer grading, discovery of novel biomarkers, recommendations for treatment plans, or evaluating the chance of recurrence. 


Our investigation focuses on the foundational tasks of classification, segmentation, and detection due to their critical importance, extensive research background, and their role in enabling further clinical inquiries. In the following, we outline various clinically recognized visual tasks, their management through DL technologies, and potential obstacles encountered by researchers in this domain. Fig. \ref{fig:apps} gives an illustration of each task.

\begin{figure*}[t]
    \centering
    \includegraphics[width=\linewidth]{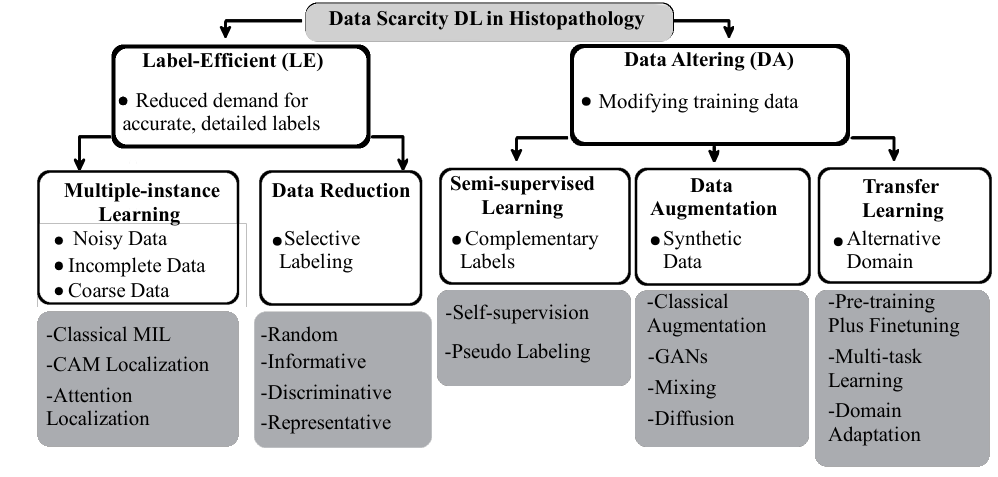}
    \caption{Summary of categories of deep learning tools addressing data scarcity in the literature of computational histopathology. Categories are indicated with boxes. The main theme in each category is indicated with bullet points.}
    \label{fig:cats}
\end{figure*}

\subsubsection{Tumor Stroma Ratio}
Studying the interactions between malignant cells and the supportive stromal environment, particularly the tumor-stroma ratio (TSR), can be highly beneficial ~\citep{TSRbreast,92, TSRcervical, TSRlung}. DL can tackle this by identifying tumor-stroma locations, classifying samples from the tumor microenvironment as tumor or non-tumor stroma, and conducting panoptic segmentation to distinguish the components of the tumor microenvironment. 

\subsubsection{Tumor Mutation Burden}

The Tumor Mutation Burden (TMB) has emerged as a crucial biomarker for predicting how well a patient might respond to immunotherapy~\citep{TMB1,bioinfo}. The standard method for measuring TMB is costly, complex, and time-intensive~\citep{TMB2}, necessitating fresh, adequately sourced tissue samples. On the other hand, initial studies have shown significant potential for DL to accurately classify TMB levels in lung adenocarcinoma~\citep{TMB2,TMB1} and CRC~\citep{TMB3} using patient WSIs, suggesting a promising direction for non-invasive, efficient cancer diagnostic and prognostic tools.

\subsubsection{Tumor-Infiltrating Lymphocytes}

Assessing the presence and concentration of Tumor-Infiltrating Lymphocytes (TILs) in a tumor provides critical insight into the body’s immune response to cancer. TILs can be distinguished in H\&E stained slides through their morphology, while accounting for important considerations. Namely, the staining color does not differentiate between lymphocyte types (e.g., T cells, B cells, NK cells), in which case some research suggests using the expensive and time-consuming IHC staining. Alternatively, DL could be particularly beneficial in identifying morphological features that may be challenging to detect by the human eye, offering an alternative to IHC testing. Furthermore, it is advised to evaluate TILs within the confines of the invasive tumor, avoiding areas affected by crush artifacts, necrosis, and inflammation near biopsy sites, and differentiating between stromal and intratumoral TILs~\citep{bioinfo2}. DL could again play a crucial role in segmenting the tumor microenvironment to precisely identify these regions.

\subsubsection{Mitotic Cells Detection and Counting}
Evaluating mitotic activity is essential for pathologists diagnosing cancer and determining its malignancy. Accurately counting mitotic figures is crucial and challenging due to the complexity of mitosis, which includes different phases, and is further complicated by the presence of lymphocytes, variability in nuclear shapes and textures, their tendency to cluster, indistinct boundaries, and changes in the microenvironment's color and texture. These challenges can be effectively handled through several DL tools geared towards object detection and segmentation.

\subsubsection{Gland Segmentation in CRC}
When detected early, colorectal cancer (CRC) can be treatable. DL technologies, especially in segmentation, are critical for automatically segmenting intestinal glands. This process extracts important quantitative features of gland morphology and structure, offering essential insights for cancer prognosis and enhancing treatment plan quality.

\subsubsection{Microsatellite Instability}
Similar to TMB, Microsatellite Instability (MSI) plays a vital role in evaluating therapeutic approaches and predicting cancer prognosis~\citep{bioinfo}, while also sharing the same challenges of being relatively expensive due to the requirements for high-quality samples and a significant presence of gene mutations, making it inaccessible for many patients. Research has shown promising results in deducing MSI status directly from WSIs of CRC and gastric adenocarcinoma, bypassing the expensive manual processes traditionally used~~\citep{MSI1,MSI3,MSI2}. 
\subsection{Survey Motivation}
\label{sec:motiv}
The efficacy of DL tools at classification, detection, and segmentation, and their great promise in the medical field have spurred extensive research into DL applications in digital pathology. Several surveys have been produced to tackle various aspects of the literature including medical image segmentation~\citep{DLsurvey, DLsegmentationSurvey}, graph-based methods for histopathology~\citep{graph-survey}, transformers in pathology image analysis~\citep{transSurvey,vit}, and DL for cancers like colon~\citep{DLcolonSurvey}, prostate~\citep{PCsurvey}, and breast~\citep{BCDsurvey2,BCDsurvey}, nucleus detection in histopathology~\citep{nucdet}, as well as identifying microsatellite instability~\citep{MSICOLsurvey}.

However, an important issue of DL application to histopathology; namely, the issue of \textbf{data scarcity}, is still in need of a detailed and expansive analysis. DL models, known for their extensive data needs, require large datasets for optimal function and effective generalization in predictive modeling. Practitioners use diverse methodologies—be it theoretical, empirical, or heuristic—to determine the necessary data volume. For instance, guidelines suggest using datasets 50 to 1000 times the number of classes~\citep{50Times}, 10 to 100 times the number of features~\citep{100Times}, or 10~\citep{10times} to 50 times~\citep{50Times2} the number of adjustable parameters in artificial neural networks (ANNs). 
 
Given the complexity of DL models, which often have numerous hidden layers and millions of parameters, assembling such datasets demands significant clinician time, exacerbating their workload and stress due to the labor-intensive data annotation process. For example, annotating 912 WSIs requires about 900 hours of clinician effort~\citep{29}, and marking 50 images, each with 12 million pixels, takes between 120 to 230 hours~\citep{59}. This scenario underscores the acute challenge of \textit{data scarcity} in medical imaging, especially for WSIs, impacting research progress and application. 

Additionally, the high cost of collecting samples, even unlabeled ones, stems primarily from the complex and expensive preparation processes involved. A critical step in WSIs is the staining procedure, such as Hematoxylin and Eosin (H\&E) or Immunohistochemistry (IHC) staining. This transforms tissues into distinguishable colors, like dark purple for nuclei and pink for other cellular structures in H\&E stains. Reports~\citep{54,64} indicate that this process can be costly, complex, time-consuming and invasive to the slide. 

Lastly, the scarcity of WSIs is compounded by privacy and ethical concerns, often restricting access to patient data for DL researchers. These factors significantly hinder the creation and development of WSI datasets \citep{challenges,ethics}.

The need to handle data-intensive modules and the challenges of acquiring such data have driven researchers to focus on enhancing data management efficiency over the past decade. Each proposed method introduces its own set of challenges and opens avenues for significant research, discovery, and innovation. Given the extensive body of research in this area, there is a recognized need for a comprehensive review that collects, organizes, and categorizes various methods, highlighting their differences, strengths, and weaknesses. Our paper addresses this by reviewing the performance of different models and methods, identifying gaps and potential improvements in the literature, and providing insights and recommendations for future research.


\begin{figure*}[t]
\begin{subfigure}[t]{0.5\textwidth}
    \centering
    \includegraphics[width=2.6in]{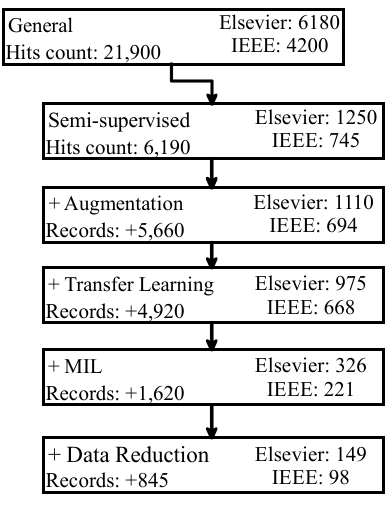}
    \caption{}
    \label{fig:searchRes-a}
\end{subfigure}
~
\begin{subfigure}[t]{0.5\textwidth}
    \centering
    \includegraphics[width=2.6in]{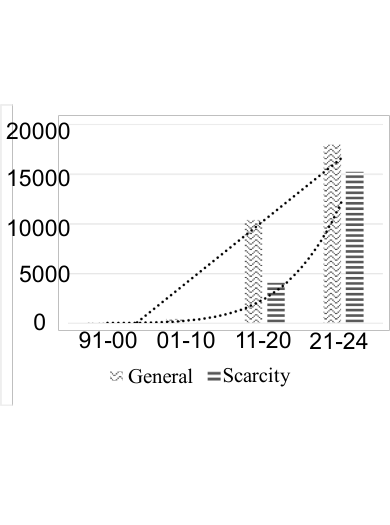}
    \caption{}
    \label{fig:searchRes-b}
\end{subfigure}
\caption{Search Results. (a) The hits count per search term. Elsevier and IEEE journals given as examples. (b) Progression of records raw count in several scientific databases. An exponential trend is observed in scarcity-oriented works in contrast to the linear trend in general works}.    
\label{fig:searchRes}    
\end{figure*}


\subsection{Data Scarcity in Histopathology}
\label{section:intro-scope}
Data scarcity manifests in various ways, depending on the specific task at hand and the methods applied. WSIs typically feature resolutions spanning millions of pixels, which poses significant computational and memory challenges; as a result, DL pipelines often start by dividing the WSI into numerous smaller sections, or patches, where the WSI is treated as a collection (``bag"), with each patch representing an element (``instance") within that collection. This is demonstrated in the upper part of Fig. \ref{fig:apps}. 

\subsubsection{Scarcity Type 1 (Few Samples)}
The classic issue of data scarcity relates to small-scale datasets that have limited annotated samples. This situation is termed Few Samples (FS) data scarcity, and it is not unique to histopathology; rather, it is commonly observed in many DL applications in the natural scene. In histopathology, FS is pronounced at patch-level predictions and describes the situation where the training set is limited in size, which causes overfitting and poor generalization to unseen phenomena. This is analogous to the Incomplete Supervision as commonly referred to in the literature \citep{weakly}.

\subsubsection{Scarcity Type 2 (Coarse Labels)}
With WSIs, it is common that the specialist evaluates the entire slide as either positive or negative for disease but does not provide detailed labels for each patch. Similarly, pathologists may provide scribbles or partial annotations in lieu of detailed and complete ones. As such, this leads to a mismatch between available data, and the requirement of DL analysis, which often involves detailed, instance-level supervision. This is commonly known as Inexact Supervision \citep{weakly}. Moreover, with WSIs, annotating the full slide and simply propagating the label to all instances causes noisy supervisory signals, as will be discussed in Section \ref{sec:mil}, which is referred to commonly as Inaccurate Supervision \citep{weakly}. Specifically, the existing annotations do not offer the level of detail and correctness needed for thorough and sound decision-making, necessitating a method to effectively bridge this gap. We refer to this scenario as Coarse Labels (CL) data scarcity.


\subsection{Data Scarcity Solutions in Histopathology}
Researchers have explored numerous strategies to overcome both types of scarcity (Few Sample and Coarse Label) in DL for histopathology. The specific tasks being addressed often dictate the choice of method, which might also introduce its own set of secondary challenges that need resolution. To organize the diverse contributions in this field effectively, we suggest classifying these strategies into two principal categories: label-efficient methods (LE) and data-altering methods (DA). Fig. \ref{fig:cats} offers a visual summary of all the discussed methods.

LE methods focus on optimizing the model, training approach, or overall training framework without altering the original training dataset. Conversely, DA methods aim to modify the training data pool. This modification enhances the use of the available labeled data alongside a wealth of unlabeled data. Although these approaches differ in their specifics, they converge on a common goal: to protect the learning model from overfitting and enhance its ability to generalize to new, unseen data, without the need to acquire additional expert-labeled data. This goal is achieved either by incorporating alternative sources of labeled data or by delving into new domains to learn useful representations (as in DA methods), or by adapting the model or training approach to be less reliant on a large volume of highly accurate, detailed labels (as in LE methods).

LE methods acknowledge the limited availability of labeled data but aim to reduce the training process's reliance on a substantial volume of highly accurate labeled data. Essentially, LE strategies operate under a limitation of resources allocated for data labeling. Within the Weakly-supervised Learning framework (Section~\ref{sec:mil}), the focus is on leveraging the existing labeled data to maximize the use of information within, despite its limited quantity and quality. Data reduction techniques (Section~\ref{sec:reduction}), consider the expertise of annotating pathologists as the key resource. The objective here is to use their knowledge effectively, while aiming to reduce the annotation workload, and navigating constraints such as limited time and budget.

On the other hand, DA techniques tackle the core issue of limited data availability. Specifically, semi-supervised learning (Section~\ref{section:semi}) incorporates unlabeled data into the training process. This is achieved either through automatic extraction of inherent labels within the data or by using a supplementary set of pseudo labels, which are simpler to acquire. Similarly, data augmentation/synthesis techniques (Section~\ref{sec:aug}) generate completely new images for inclusion in the learner's training process. In contrast, transfer learning approaches (Section~\ref{section:TL}) use datasets that, while different, are closely related, incorporating their labels and tasks into the training.

\subsection{Survey Scope}

Existing reviews addressing data scarcity in DL applications to CPath focus on several important aspects. However, to the best of our knowledge, there is still a need for a comprehensive review that consolidates all existing approaches to data scarcity, as most studies tend to selectively focus on specific techniques, such as semi-supervised learning or a combination of semi-supervised and transfer learning \citep{notso,strategies,milReview,survey-class,survey2,survey}. This selective focus makes it difficult to discern which techniques are thoroughly examined and which remain underexplored, creating gaps in understanding and limiting the identification of new research opportunities. In this survey, we aim to address these gaps by offering a more holistic and consolidated review. Additionally, we have included a dedicated section (Section \ref{sec:reduction} - Data Reduction) that covers a family of methods not explicitly discussed in prior reviews. Our findings suggest that these methods, despite their significant potential, are under-explored. In Table. \ref{tab:existing}, we compare our review with existing ones, highlighting areas that we have comprehensively addressed. This comparison is intended to guide future research by benchmarking current methods and offering recommendations for further exploration, providing a well-rounded perspective on the challenges of data scarcity in CPath.


\subsection{Article Search Criteria}
Our literature review, focusing on digital pathology and deep learning applications, employs the search phrase ``(WSI OR histopathology OR ``whole-slide images") AND ``Deep Learning" AND (Classification OR Segmentation OR Detection)" to obtain an overview of research targeting the three primary tasks in digital pathology from 2015 to 2024. We include all relevant studies without filtering for their impact, yet we conduct an in-depth review within the main databases, concentrating on significant works only. To address the issue of data scarcity, we apply an exclusion criterion using the ``AND" operator, selecting for scarcity-related research with terms like (scarcity OR ``limited data" OR ``label efficient" OR ``small scale" OR weak). We refine the search with additional ``OR" operations for specific categories previously outlined. All such categories and their search terms are summarized in Table.\ref{tab:search}.

Our findings are summarized in Fig. \ref{fig:searchRes-a}, which reveals that semi-supervised methods predominate, with augmentation and transfer learning approaches closely behind. Data reduction is the least represented category. A temporal trend analysis in Fig. \ref{fig:searchRes-b} expands the search to include the evolution of general versus scarcity-focused research over time, indicating an overall increase in publications. The past four years, starting in 2021, have seen a surge in publications, outpacing the entire previous three decades. This trend is anticipated to continue through the end of 2030, for both general and scarcity-specific studies. Notably, the gap between general and scarcity-oriented research is narrowing, as depicted by the declining ratio of general to scarcity-focused publications (from 10.4 down to 1.3), and also evidenced by the exponential increase in scarcity-oriented work, underscoring the escalating focus by the scientific community on this issue. 

\begin{table}[]
\centering

\caption{Article search criteria to query different aspects of the review.}
\label{tab:search}
\begin{tabular*}{3in}{>{\footnotesize}l>{\footnotesize}l}
\toprule
\multicolumn{1}{>{\footnotesize}c}{\textbf{Search Term}} & \multicolumn{1}{>{\footnotesize}c}{\textbf{Description}} \\ \midrule
\begin{tabular}[c]{@{}l@{}}(WSI OR histopathology \\ OR ``whole-slide images") AND \\ ``Deep Learning" AND\\  (Classification OR Segmentation OR\\  OR Detection)\end{tabular} & \begin{tabular}[c]{@{}l@{}}All general works on DL \\ applied to Histopathology, \\ concentrating on the\\ three main tasks\end{tabular} \\ 
\addlinespace
\begin{tabular}[c]{@{}l@{}}scarcity OR ``limited data" OR\\  ``label efficient" OR ``small scale" OR\\ weak\end{tabular} & Addressing data scarcity \\ 
\addlinespace
\begin{tabular}[c]{@{}l@{}}``MIL" OR \\ ``multiple instance learning"\end{tabular} & Multiple-instance Learning \\ 
\addlinespace
\begin{tabular}[c]{@{}l@{}}``data reduction" OR \\ ``selective labeling" OR \\ ``active learning"\end{tabular} & Data Reduction \\ 
\addlinespace
\begin{tabular}[c]{@{}l@{}}semi OR self OR\\ ``contrastive learning" OR \\ ``pseudo label" OR pretext\end{tabular} & Semi-supervised Learning \\ 
\addlinespace
\begin{tabular}[c]{@{}l@{}}augmentation OR synthesis OR \\ ``synthetic images"\end{tabular} & Data Augmentation \\ 
\addlinespace
\begin{tabular}[c]{@{}l@{}}``transfer learning" OR \\ ``domain adaptation" OR \\ ``multi-task learning"\end{tabular} & Transfer Learning \\
\bottomrule
\end{tabular*}
\end{table}


 \subsection{Benchmarking}
Our review meticulously catalogs the existing literature, presenting tables that benchmark performances across pivotal histopathology tasks — classification, detection, and segmentation. We examine classifiers by assessing their accuracy (ACC), the area under the receiver operating characteristic curve (AUC)~\citep{auc}, as well as their Kappa~\citep{kappa} and F1 scores. Segmentors and detectors are evaluated on the basis of the Dice coefficient (D) and the Jaccard Index (J). For data augmentation techniques, we further include quality metrics such as the Frechet Inception Distance (FID)~\citep{fid}, the Structural Similarity Index (SSI)~\citep{ssim}, Feature Similarity Index (FSI)~\citep{fsim}, the Gradient Magnitude Similarity Deviation (GMSD)~\citep{gmsd}, and the Normalized Root Mean Square Error (NRMSE). Through these multifaceted evaluation protocols, we aim to provide a nuanced analysis of each method's performance in addressing the core tasks of digital pathology.

\begin{table*}[]
\centering
\caption{Comparison of our review with other reviews that address data scarcity in CPath. We aim to maximize the utility of our review by exhaustively addressing all relevant aspects in the art. Localization (Loc.), in contrast to classification (CLS) alone, is of special interest in CPath.}
\label{tab:existing}
\begin{tabular}{>{\footnotesize}l>{\footnotesize}c>{\footnotesize}c>{\footnotesize}c>{\footnotesize}c>{\footnotesize}c>{\footnotesize}c}
\toprule
\textbf{Survey} & \citep{strategies} & \citep{survey2} & \citep{survey} & \citep{survey-class} & \citep{milReview} & Ours \\ \midrule
\textbf{MIL} & \checkmark & \checkmark & \checkmark & \xmark & \checkmark & \checkmark \\ 
\textbf{Data Reduction} & \xmark & \xmark & \xmark & \xmark & \xmark & \checkmark \\ 
\textbf{\begin{tabular}[c]{@{}l@{}}Semi-supervised\\ Learning\end{tabular}} & \checkmark & \xmark & \checkmark & \checkmark & \xmark & \checkmark \\ \addlinespace
\textbf{\begin{tabular}[c]{@{}l@{}}Data \\ Augmentation \end{tabular}} & \checkmark & \checkmark & \xmark & \xmark & \checkmark & \checkmark \\ \addlinespace
\textbf{\begin{tabular}[c]{@{}l@{}}Transfer\\ Learning\end{tabular}} & \checkmark & \checkmark & \xmark & \xmark & \xmark & \checkmark \\ \addlinespace
\textbf{CLS + Loc.} & \xmark & \checkmark & \checkmark & \xmark & \xmark & \checkmark \\ 
\textbf{\begin{tabular}[c]{@{}l@{}}Bench-\\ marking\end{tabular}} & \xmark & \xmark & \checkmark & \checkmark & \xmark & \checkmark \\ \addlinespace
\textbf{\begin{tabular}[c]{@{}l@{}}Future\\ Directions\end{tabular}} & \xmark & \checkmark & \checkmark & \checkmark & \xmark & \checkmark \\
\bottomrule

\end{tabular}
\end{table*}

Methods that address data scarcity in digital pathology are assessed by examining their impact on a learner's performance on histopathology tasks; namely, by comparing results obtained before and after applying the data scarcity solution. To gauge this effectively, our review provides a ``performance improvement/retention percentage" (Perf.\%) which is robust and consistent across the different approaches. Additionally, we indicate in a specific column the percentage of data utilized by the proposed method compared to the data amount used to train the baseline. The merit of a method is evident if it can either diminish the need for annotations while only minimally reducing baseline performance or if it can enhance baseline performance without increasing the need for annotations. For consistency in our evaluation, we average the best scores reported in cases where multiple outcomes are provided within an ablation study and intend to report 40x magnification results accross all works, and commonly at 20\% of the data volume simulating the data scarcity setting.


\subsection{Survey Contributions}

On the whole, the distinct contributions of this review paper can be summarized by the following points:
\begin{itemize}
    \item We provide a structured index of research on data scarcity over the past decade, categorizing works in a way that enhances understanding and facilitates comparisons.
    \item We propose a novel and logical categorization of techniques which separates Label-efficient from Data-altering methods.
    \item We highlight the strengths, shortcomings, and most relevant challenges faced by researchers, as well as potential research gaps, and chronological trends in the literature to guide future research studies. 
    \item Our review addresses gaps in previous analyses by incorporating a comprehensive spectrum of remedies for data scarcity. We delve into an under-explored research area, providing a detailed context within the broader literature.
    \item We present detailed benchmarking tables that quantitatively compare various solutions, highlighting their effectiveness and limitations.  
    \item We introduce `Performance Improvement/Retention Percentage' (Perf.\%), a robust metric designed to assess the impact of each remedy on data efficiency.
\end{itemize}


\section{Label Efficient (LE) Methods}
In this section, we discuss the Label-Efficient strategies, which aim to optimize the performance of DL models when detailed labeling is sparse. We focus on two principal categories within this approach: Multiple Instance Learning (MIL) and Data Reduction. Each category demonstrates how effective learning can be achieved with minimal direct annotation, thus reducing the workload and dependency on extensive labeled datasets.


\subsection{Multiple Instance Learning (MIL)} 
\label{sec:mil}

\begin{figure*}[t]
    \centering
    \begin{subfigure}{5.3 in}
        \includegraphics[width = 5.3in]{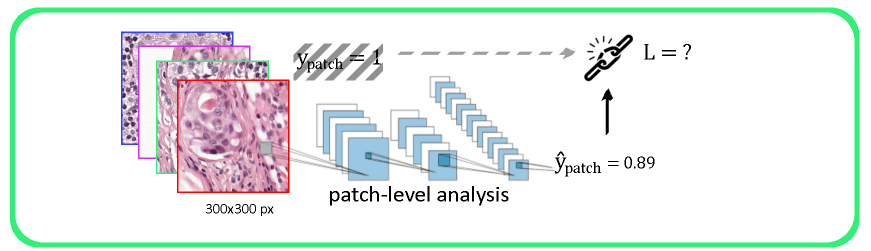}
        \caption{}  
        \label{fig:weak-cls}
    \end{subfigure}
    \begin{subfigure}{5.3 in}
        \includegraphics[width = 5.3 in]{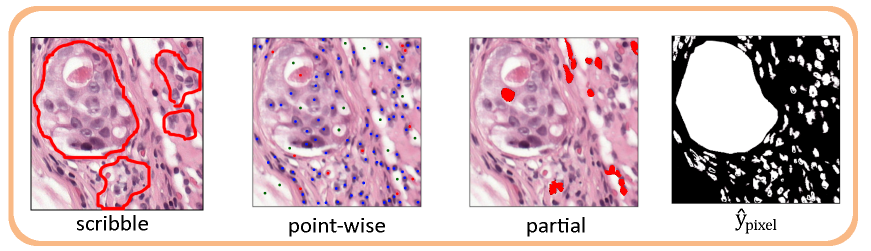}
        \caption{} 
        \label{fig:weak-seg}
    \end{subfigure}
    
    \caption{Illustration of weak supervisory signals handled within the MIL setting in WSIs: In (a), the classification case involves analysis on the patch level, producing $\hat{y}_{patch}$, whereas only slide label $Y$ is given, and the real $y_{patch}$ is inaccessible or inaccurate. In (b), the segmentation task involves pixel values prediction $\hat{y}_{pixel}$, while the given annotation is rough or partial.}
    
\end{figure*}

Methods within the MIL framework often address the aforementioned Coarse Labels (CL) type of data scarcity, which involves dealing with weak supervisory signals. Specifically, MIL strategies are geared towards leveraging valuable insights from samples that are assigned at a general level, and which may additionally produce noisy information—conditions that are generally less demanding than acquiring precise and clean labels. 

Consider a WSI, divided into $N$ patches $x_i$ within the set $X$, where $i \in [1-N]$. Let $Y \in \{0,1\}$ represent the label for the WSI $X$, and $y_i$ denote the label for each patch $x_i$. In the scenario characterized by CL type data scarcity, the overall label $Y$ is known, but the individual labels $y_i$ for each patch are not. This is illustrated in Fig.~\ref{fig:weak-cls}. The fundamental assumption of MIL posits that if $Y=0$, then it follows that the sum of all $y_i$ equals $0$. This assumption, while logical, leaves open questions regarding the identification of specific instance labels when $Y=1$. Namely, it is unclear which $y_i$ instances are indicative of this outcome.

Furthermore, the instances $(x_i,y_i)$ may represent pixel or region instances within a patch, where 
$(X,Y)$ pertains to a single patch. In this scenario, a weak supervisory signal indicates the class label of the entire patch, with the task at hand being the localization of histopathological objects or achieving pixel-wise segmentation. Other forms of weak supervisory signals include scribbles, point-wise, and partial annotations, particularly when addressing tasks such as pixel-wise segmentation. These concepts are visually depicted in Fig. \ref{fig:weak-seg}.

Essentially, MIL aims to enhance the efficacy of DL tools despite the discrepancies between the level of detail in provided annotations and the granularity required by the task at hand, typically through Equation.\ref{eq:MIL}, where $\mathscr{F}$ represents an instance-level feature extractor, and $\mathscr{H}$ denotes an aggregation network.

\begin{equation}
    z_i = \mathscr{F}(x_i),
    \hat{Y} = \mathscr{H}(z_i)
    \label{eq:MIL}
\end{equation}


\subsubsection{Types of MIL}
MIL approaches can be categorized into two primary types: decision-fusion MIL and feature-fusion MIL. {\bf Decision-fusion MIL}, often referred to as instance-level MIL, involves assigning the overall bag label $Y$ to its constituent instances $y_i$ , thereby allowing for the training of a classifier on each individual instance. The predictions from these instance-level classifiers are then combined to produce a prediction for the entire bag. In this framework, the $\mathscr{F}$ module generates instance-level predictions $z_i = \hat{y}_i$, and the $\mathscr{H}$ aggregator focuses on combining these decisions rather than merging features.
{\bf Feature-fusion MIL}, on the other hand, known as bag-level MIL, involves the aggregation of extracted instance features $z_i$ into a comprehensive bag-level representation. A classification decision is then made based on this aggregated feature set. The process of both aggregation and classification may be conducted by the $\mathscr{H}$ module alone or be divided among two distinct modules, as seen in certain studies. The primary goal of this approach is to achieve precise WSI classification $\hat{Y}$, with an emphasis on interpretability via localization.

%


\subsubsection{Challenges in MIL}
One of the primary challenges faced by decision-fusion methodologies is the potential disconnect between the overall label of a bag ($Y$) and the actual, but unknown, labels of its individual elements ($y_i$). Typically, a WSI contains a relatively small proportion of instances that actually carry the disease. These are the instances where  $Y$ is indicated by $y_i$, whereas the vast majority of instances may not show any signs of the disease. The fundamental assumption in MIL offers limited insight into the labels of individual instances when $Y = 1$. Consequently, training a model at the instance level can result in receiving a noisy supervisory signal. This necessitates a careful approach to the subsequent fusion of decisions. Moreover, it s crucial to define an appropriate framework for aggregating the predictions of different patches (in the case of decision-fusion MIL) or their features (in feature-fusion MIL). The development of such a framework has captured the interest of numerous researchers. It is widely accepted that the key to enhancing classifier performance lies in both, establishing an effective aggregation method and learning superior features.

It is also important to highlight that devising a bag classifier, based on the collective feature representation of all instances (as done in feature-fusion MIL), tends to obscure the location-specific information, focusing instead on the classification of the bag as a whole. Empirical evidence suggests that the most effective bag classifier may not perform equally well at classifying individual instances \citep{mil-stability}. This observation underscores the necessity for exercising caution with feature-fusion methods to maintain their interpretability. In contrast, decision-fusion approaches, along with detection and segmentation techniques, inherently provide interpretable outcomes due to their localized framework.

It should be noted that studies opting to bypass the MIL framework entirely~\citep{notMIL,notMIL2}, aiming instead to reduce computational and memory demands for a direct end-to-end classification of WSIs, are generally outside the scope of this survey. The exception to this would be those offering visualization techniques based on Class Activation Maps (CAM)~\citep{137} or similar methods.

\subsubsection{Classical Paradigm} 
Historical developments in histopathology primarily focused on improving the universality and expressiveness of the features extracted, automating the process of feature extraction, and enhancing the speed of the decision-making phase. In this context, the work of ~\citet{168} is an early advancement in applying DL-based MIL to histopathology. The aim was to automate the extraction of features, mirroring the contemporary shift from traditional ML to DL approaches. Their methodology prominently makes use of Convolutional Neural Networks (CNNs) for extracting features and focuses on analyzing patch-level instances. The attempt to automate feature extraction was further pursued in subsequent studies~\citep{166,27,169}. Specifically, \citet{169} employ CNNs to extract features from WSIs, albeit with a pre-training on natural images, which may not be ideal due to significant differences in domains as elaborated in Section~\ref{section:TL}. At the same time, the general MIL community favored the use of average-pooling and max-pooling for aggregating instance labels~\citep{concurrent}.

\begin{figure*}[t]
    \centering
    \includegraphics[width=\linewidth]{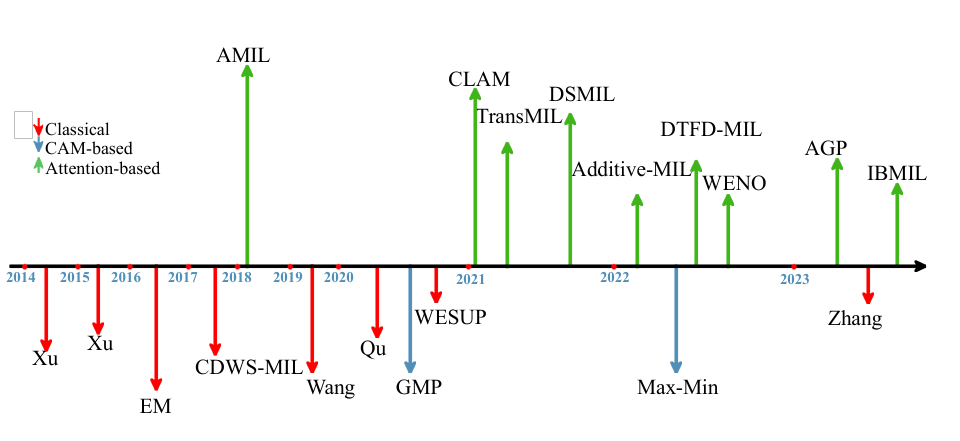}
    \caption{Chronological Trend Line of the Most Relevant MIL-based Techniques. The length of an arrow is proportional to its popularity per its category. Most classical methods are published before 2020, whereas most recent works are attention-based.}
    \label{fig:weak_trend}
\end{figure*}

Subsequent research went beyond merely automating feature extraction with CNNs, aiming to derive more valuable insights from the labeled images available. Hou \textit{et al.}\citep{166} proposed a method combining patch-level CNNs with an Expectation Maximization (EM) strategy to assess the likelihood of patch-level predictions, thus probing the distinctiveness of each patch. This approach uses an iterative process to progressively include more distinctive patches in the training set, relying on previous predictions for labeling. Similar methods have since emerged~~\citep{140}, which focus on iteratively extracting and incorporating high-confidence predictions from regions of interest, thus improving segmentation performance. These techniques align closely with Pseudo Labeling methods (Section \ref{section:semi-pseudo}), although with the aim to extract distinctive features rather than expand the training dataset. \citet{21} introduced the Constrained Deep Weak Supervision MIL (CDWS-MIL) for segmentation, using a pixel-level classifier that aggregates predictions into an image-level forecast with a Generalized Mean softmax function and compares this to ground truth image-level annotations. A critical feature of this method is that it imposes constraints on the estimated size of cancerous areas, leveraging minimal annotations. Despite relying on weak supervisory signals, CDWS-MIL has achieved results comparable to fully-supervised baselines.

\citet{17} developed a technique for pixel-wise nuclei segmentation using partially, point-wise annotated nuclei. They utilized extended Gaussian masks derived from the labels to train a U-Net model. The segmentation was refined through Voronoi segmentation and k-means clustering to separate nuclei from the background. This training was iterative, progressively incorporating new annotation points, achieving near-full performance with only 25\% of the nuclei annotated, as shown in Table.~\ref{tab:bench-mil}. Similarly, the weakly-supervised (WESUP) method of \citet{13} segments cancerous tissues using partial, point-wise annotations. This approach uses the Simple Linear Iterative Clustering (SLIC) to analyze images at the superpixel level. Superpixels are labeled based on a majority vote of sparsely labeled points or adopt the label of their closest superpixels, defined by deep hierarchical features. Despite training with only 40 sparsely annotated points per patch, this method outperforms fully-supervised baselines, as indicated in Table.~\ref{tab:bench-mil}.

Recent research by \citet{158} aims to improve feature extraction from histopathology images by minimizing the influence of bag contextual priors, which can distort the label-image relationship. Their Interventional Bag MIL (IBMIL) strategy identifies confounders from feature extractor's high-level layers, compiling them into a diverse bag-level feature dictionary. Bags are clustered in feature space, and a collective cluster-level feature is used as a dictionary unit. This method, which enhances many existing techniques, showed improved classification performance compared to AMIL, as seen in Table.~\ref{tab:bench-mil}. \citet{40} used k-means clustering on features from a Tiny-ViT module to identify key instances, which are further processed for enhanced embedding. They also established a knowledge distillation process between streams trained on lymph and thyroid data, improving feature quality and overall model performance.

Wang and peers \citep{27} proposed a feature-fusion MIL aiming to reduce the time overhead of decision-fusion approaches~\citep{168,166} in addition to automating feature extraction. The authors incorporated an additional supporting weak supervisory signal using rough polygonal segmentations of tumor regions in some samples, giving greater weight to annotated areas. Moreover, feature extraction is applied to the most reliable and distinctive regions, or blocks within the WSIs for more efficient and accurate feature extraction and aggregation. \citet{152} employ partial annotation, specifically making use of in-target and out-of-target annotation points (located within and outside the target-of-interest, respectively). By this approach, the authors aim to delineate the distinctly contrasting features of in-target and out-of-target points. 


\subsubsection{CAM-based Localization}
In addition to accelerating inference, methods that merge features in the MIL setting (feature-fusion MIL methods) have outperformed those that aggregate decisions (decision-fusion methods) in classifying WSIs~\citep{bagBetter}. Consequently, more contemporary approaches are adopting a framework that analyzes at the bag level, while carefully addressing their shortcomings in localization. \citet{137} developed a method to classify WSIs directly, bypassing the traditional MIL frameworks. They introduced the Global Mean Pooling (GMP) technique, which leverages CUDA's Unified Memory (UM) system to efficiently transfer tensors from GPU memory, incorporating optimization strategies to minimize data transfer delays. Additionally, they utilized Class Activation Mapping (CAM)-based visualizations \citep{main-cam} to produce heatmaps post-classification, highlighting histopathological features relevant to the predictive class using only the slide-level label as a weak supervisory signal. In related work, \citet{19} also employed CAM-based localization to distinguish between regions of high confidence, which positively influence predictions, and regions of high uncertainty, which do not, using a Kullback-Leibler (KL) divergence term \citep{kl}. Their goal was to clearly separate foreground from background in histopathology images, often visually similar, to reduce false positives and enhance both classification and localization accuracy, improving upon traditional Grad-CAM methods, as detailed in Table.~\ref{tab:bench-mil}.


\subsubsection{Attention-based Localization}

The Attention-based Multiple Instance Learning (AMIL) framework, introduced by \citet{AMIL}, represents a significant advancement in the art. It aims to improve the interpretability of feature-fusion MIL by pinpointing specific instances within a bag that significantly influence the overall bag label. This is achieved through a parametric, trainable fusion mechanism for instance embeddings, utilizing a two-layer neural network that implements a gated attention mechanism \citep{bahdanau}. Unlike traditional static average pooling methods~\citep{concurrent}, this approach provides a dynamic, learned weighted average where weights are determined by the attention between the bag label and individual instances. This enables the estimation of instance-level labels, which are typically unavailable in feature-fusion methods. AMIL has demonstrated improved classification accuracy over traditional methods that use average pooling, as highlighted in Table.~\ref{tab:bench-mil}.

\begin{table*}[]
\centering
\caption{Summary of the most relevant MIL approaches per type, organized chronologically in descending order. Repeated datasets are in bold. Performance \% is w.r.t the baseline method mentioned in the work. Localization support is indicated by a checkmark.}
\label{tab:bench-mil}
\begin{tabular}{>{\footnotesize}l>{\footnotesize}c>{\footnotesize}c>{\footnotesize}c>{\footnotesize}c>{\footnotesize}c}
\toprule
\multicolumn{1}{>{\footnotesize}c}{\textbf{Type}} & \textbf{Work} & \textbf{Task} & \textbf{Dataset} & \textbf{Loc.} & \textbf{Perf.\%} \\ \midrule
\multirow{22}{*}{\rotatebox[origin=c]{45}{Attention-based}} & \multirow{2}{*}{IBMIL - \citep{158}} & \multirow{2}{*}{CLS} & \textbf{Camelyon16 - Breast} & \multirow{2}{*}{\checkmark} & \textcolor{green}{$\uparrow$}3.42\% (ACC) \textcolor{green}{$\uparrow$}1.62\% (AUC) \\
 &  &  & \textbf{TCGA - Lung} &  & \textcolor{green}{$\uparrow$}1.57\% (ACC) \textcolor{green}{$\uparrow$}1.08\% (AUC) \\ \addlinespace 
 & \multirow{2}{*}{AGP - \citep{164}} & \multirow{2}{*}{CLS} & SICAPv2 - Prostate & \multirow{2}{*}{\checkmark} & \textcolor{green}{$\uparrow$}5.88\% (k) \\
 &  &  & PANDA - Prostate &  & \textcolor{green}{$\uparrow$}1.74\% (k) \\ \addlinespace 
 & \multirow{3}{*}{WENO - \citep{153}} & \multirow{3}{*}{CLS} & Private - Cervical & \multirow{3}{*}{\checkmark} & \textcolor{green}{$\uparrow$}24.97\% (AUC) \\
 &  &  & \textbf{Camelyon16 - Breast} &  & \textcolor{green}{$\uparrow$}3.38\% (AUC) \\
 &  &  & \textbf{TCGA - Lung} &  & \textcolor{green}{$\uparrow$}1.84\% (AUC) \\ \addlinespace 
 & \multirow{3}{*}{Additive-MIL - \citep{150}} & \multirow{3}{*}{CLS} & \textbf{Camelyon16 - Breast} & \multirow{3}{*}{\checkmark} & \textcolor{green}{$\uparrow$}12.8\% (AUC) \textcolor{green}{$\uparrow$}7.37\% (ACC) \\
 &  &  & \textbf{TCGA - Lung} &  & \textcolor{red}{$\downarrow$}0.53\% (AUC) \textcolor{green}{$\uparrow$}0.34\% (ACC) \\
 &  &  & \textbf{TCGA - Kidney} &  & \textcolor{green}{$\uparrow$}0.51\% (AUC)\textcolor{green}{$\uparrow$}4.21\% (ACC) \\ \addlinespace 
 & \multirow{2}{*}{DTFD-MIL - \citep{24}} & \multirow{2}{*}{CLS} & \textbf{Camelyon16 - Breast} & \multirow{2}{*}{\checkmark} & \begin{tabular}[c]{@{}c@{}}\textcolor{green}{$\uparrow$}10.77\% (AUC) \textcolor{green}{$\uparrow$}7.45\% (ACC)\\ \textcolor{green}{$\uparrow$}13.08\% (F)\end{tabular} \\
 &  &  & \textbf{TCGA - Lung} &  & \textcolor{green}{$\uparrow$}2.13\% (AUC)\textcolor{green}{$\uparrow$}2.88\% (ACC) \\ \addlinespace 
 & \multirow{2}{*}{DSMIL - \citep{34}} & \multirow{2}{*}{CLS} & \textbf{TCGA - Lung} & \multirow{2}{*}{\checkmark} & \textcolor{green}{$\uparrow$}1\% (AUC) \textcolor{green}{$\uparrow$}3.18\% (ACC) \\
 &  &  & \textbf{Camelyon16 - Breast} &  & \textcolor{green}{$\uparrow$}3.3\% (AUC) \textcolor{green}{$\uparrow$}2.65\% (ACC) \\ \addlinespace 
 & \multirow{3}{*}{TransMIL - \citep{171}} & \multirow{3}{*}{CLS} & \textbf{Camelyon16 - Breast} & \multirow{3}{*}{\checkmark} & \textcolor{green}{$\uparrow$}6.27\% (AUC) \textcolor{green}{$\uparrow$}1.79\% (ACC) \\
 &  &  & \textbf{TCGA - Lung} &  & \textcolor{green}{$\uparrow$}10.94\% (AUC) \textcolor{green}{$\uparrow$}14.46\% (ACC) \\
 &  &  & \textbf{TCGA - Kidney} &  & \textcolor{green}{$\uparrow$}1.86\% (AUC) \textcolor{green}{$\uparrow$}5.95\% (ACC) \\ \addlinespace 
 & \multirow{3}{*}{CLAM - \citep{170}} & \multirow{3}{*}{CLS} & \textbf{Camelyon16/17 - Breast} & \multirow{3}{*}{\checkmark} & \textcolor{red}{$\downarrow$}4.7\% (AUC)* \\
 &  &  & \textbf{TCGA - Lung} &  & \textcolor{red}{$\downarrow$}4.4\% (AUC)* \\
 &  &  & \textbf{TCGA - Kidney} &  & \textcolor{red}{$\downarrow$}0.9\% (AUC)* \\ \addlinespace 
 & \multirow{2}{*}{AMIL - \citep{AMIL}} & \multirow{2}{*}{CLS} & Colon Cancer & \multirow{2}{*}{\checkmark} & \textcolor{green}{$\uparrow$}2.98\% (AUC) \textcolor{green}{$\uparrow$}5.12\% (ACC) \\
 &  &  & Breast Cancer &  & \textcolor{green}{$\uparrow$}0.38\% (AUC) \textcolor{green}{$\uparrow$}1.89\% (ACC) \\ \midrule
\multirow{4}{*}{\rotatebox[origin=c]{45}{CAM-based}} & \multirow{2}{*}{Max-Min Uncertainty - \citep{19}} & \multirow{2}{*}{CLS} & \textbf{Camelyon16 - Breast} & \multirow{2}{*}{\checkmark} & \textcolor{green}{$\uparrow$}8.73\% (D) \\
 &  &  & GlaS - Colorectal &  & \textcolor{green}{$\uparrow$}8.76\% (D) \\ \addlinespace 
 & \multirow{2}{*}{GMP - \citep{137}} & \multirow{2}{*}{CLS} & \textbf{TCGA - Lung} & \multirow{2}{*}{\checkmark} & \textcolor{green}{$\uparrow$}4.06\% (AUC) \\
 &  &  & TMUH, WFH, and SHH - Lung &  & \textcolor{green}{$\uparrow$}3.05\% (AUC) \\ \midrule
\multirow{9}{*}{\rotatebox[origin=c]{45}{Classical}} & \multirow{2}{*}{WESUP - \citep{13}} & \multirow{2}{*}{SEG} & CRAG - Colorectal & \multirow{2}{*}{\checkmark} & \textcolor{green}{$\uparrow$}0.37\% (D) \\
 &  &  & \textbf{GlaS - Colorectal} &  & \textcolor{green}{$\uparrow$}1.17\% (D) \\ \addlinespace 
 & \multirow{2}{*}{Qu - \citep{17}} & \multirow{2}{*}{SEG} & MoNu - Multi-organ & \multirow{2}{*}{\checkmark} & \textcolor{red}{$\downarrow$}7.27\% (D) \\
 &  &  & Private - Lung &  & \textcolor{red}{$\downarrow$}11.44\% (D) \\ \addlinespace 
 & Wang - \citep{27} & CLS & Private - Lung & \multicolumn{1}{l}{} & \textcolor{green}{$\uparrow$}4.62\% (ACC) \\ \addlinespace 
 & \multirow{2}{*}{CDWS-MIL - \citep{21}} & \multirow{2}{*}{SEG} & randomly selected from \citep{21-related-dataset} & \multirow{2}{*}{\checkmark} & \textcolor{red}{$\downarrow$}8.39\% (D) \\
 &  &  & Private - Colon &  & \textcolor{red}{$\downarrow$}4.57\% (D) \\ \addlinespace 
 & \multirow{2}{*}{EM - \citep{166}} & \multirow{2}{*}{CLS} & \textbf{TCGA - Lung} & \multicolumn{1}{l}{\multirow{2}{*}{}} & \textcolor{green}{$\uparrow$}8.13\% (ACC) \\
 &  &  & TCGA - Brain Glioma & \multicolumn{1}{l}{} & \textcolor{red}{$\downarrow$}0.24\% (ACC) \\
 \bottomrule
\end{tabular}
\end{table*}

Building on the foundation laid by AMIL, the Clustering-Constrained Attention MIL (CLAM) model proposed by \citet{170} enhances interpretability within the MIL framework through a novel multi-class classification approach. CLAM introduces multiple attention branches, corresponding to the number of classes (N), each leading to a distinct classification head, providing separate slide-level representations for each class. \citet{171} introduced TransMIL, which integrates the Transformer module to model the attention between instances based on a self-attention operation. This method captures both local and global representations, improving accuracy and accelerating model convergence.

Additionally, the Dual-Stream MIL (DSMIL) model by \citet{34} utilizes a dual-faceted strategy to focus on a critical instance, enhancing attention relations with other instances. DSMIL integrates a contrastive learning strategy inspired by SimCLR \citep{simclr} and extracts features at various magnifications to form a comprehensive multi-scale embedding, mirroring advancements seen in integrated CNN approaches at both image and patch levels~\citep{23}. Zhang and others modify AMIL by employing grad-CAM over the attention path to derive instance-level labels within a feature fusion framework~\citep{24}. They introduced the double-tier feature distillation multiple instance learning (DTFD-MIL), which condenses data collections (``bags") into smaller ``pseudo bags" for feature extraction at both pseudo and parent bag levels, outperforming the AMIL framework as shown in Table. ~\ref{tab:bench-mil}.

Addressing some limitations of attention-based localization methods, \citet{150} identified challenges such as distinguishing between positively and negatively attended instances and the general insensitivity to the specific class being predicted. They proposed an Additive MIL approach that tracks each instance’s contribution to the overall prediction, modifying the traditional feature averaging method of AMIL to be the sum of contributions from each instance. Furthermore,\citet{153} developed a weakly-supervised knowledge distillation framework named WENO, using a teacher-student model. The teacher model, trained to perform slide classification, generates attention-based heatmaps as pseudo-labels for a student model trained to perform patch-level classification. Moreover, the student distills knowledge to the teacher by sharing instance feature extractors with the teacher, thus establishing bi-directional knowledge transfer, leading both models' performance to improve during training. Their results are summarized in Table.\ref{tab:bench-mil}.

\citet{164} introduced Attention Gaussian Processes (AGP); an inherently nondeterministic approach, to provide a probabilistic approximation for AMIL over instance-level labels, thus aligning with typical clinical practices where pathologists often include a margin of uncertainty in their diagnoses. The approach has shown improvements in slide-level classification of prostate cancer images compared to the baseline mean aggregation method, as demonstrated in Table.\ref{tab:bench-mil}.

\begin{figure*}[t]
    \centering
    \includegraphics[]{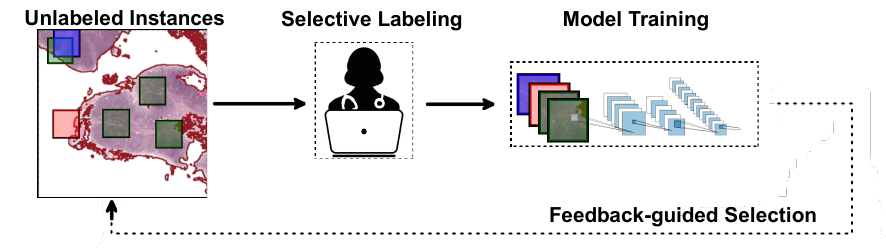}
    \caption{Illustration of the Data Reduction Paradigm. An exclusive selection of instances are relayed to an expert for annotation. Red samples indicate non-informative instances, blue samples indicate duplicated phenomena, and green samples indicate informative and representative instances. The dashed arrow represents some non-random selection methods where the query is informed by the learner's behavior.}
    \label{fig:reduction}
\end{figure*}
\subsubsection{Summary}
Datasets such as TCGA~\citep{tcga} and Camelyon16~\citep{cam16} are frequently used in MIL research, as outlined in Table.~\ref{tab:bench-mil}, which compares results from various MIL approaches across different datasets. Fig.~\ref{fig:weak_trend} provides a chronological overview of these methodologies. Key observations from MIL techniques are also highlighted, emphasizing their application and trends:


\begin{itemize}
    \item 
    Typically, MIL approaches address data scarcity of the Coarse Labeling (CL) type, where a limited annotation budget results in labels that are less detailed than the analytical or inferential resolution required.
    
    \item 
    There is often a degree of flexibility in the application of MIL definitions across studies, including practices such as acquiring instance labels for a subset of instances, working with approximately-annotated areas, images with partial annotations, or point annotations, as well as incorporating specific priors into the analysis.
    
    \item 
    The use of MIL extends beyond classification tasks to include segmentation and detection tasks. In these cases, image patches are treated as ``bags" containing individual pixels as ``instances".
    
    \item 
    Common trends observed in MIL research include identifying discriminative patches, automating feature extraction, adopting a phased, iterative approach to training, and efforts to speed up the inference process.
    
    \item 
    The challenge of segmentation tasks and the degree of supervision varies. For example, the study in~\citep{17} uses sparse, point-wise annotations, which represent a lesser degree of supervision compared to the approach in~\citep{140}.
    
    \item 
    Recent developments in MIL predominantly focus on localization issues, highlighting the growing emphasis on interpretability within DL tools as a prerequisite for clinical implementation.
    
    \item 
    Following the introduction of Attention-based Multiple Instance Learning (AMIL)~\citep{AMIL} in 2018, most subsequent methodologies have adopted or modified the AMIL framework. Conversely, the integration of Class Activation Mapping (CAM)-based visualizations has not been a primary focus in most of these contributions.
    
\end{itemize}


\subsection{Data Reduction}
~
\label{sec:reduction}
Data reduction, also referred to as ``sample-selection" or ``selective labeling", represents an alternative strategy for addressing data scarcity while maximizing label efficiency. This approach prioritizes the strategic annotation of data by choosing a subset of the dataset for expert annotation. The selection aims to preserve the dataset's informative value, while lessening the annotation workload. Beyond lowering annotation costs, these methods focus the limited annotation resources on the most informative samples, potentially leading to higher quality annotations compared to scribbles or point-wise annotations. Furthermore, data reduction techniques can significantly benefit clinical practices by highlighting informative and crucial areas thus enhancing the interpretability of DL tools, and fostering novel clinical discoveries.

In addition to the selection mechanism, it is common in this domain to employ an Active Learning (AL) framework to incrementally include chosen samples into the training set. This process is repeated until a saturation point in accuracy is achieved or available resources are fully used. Studies have demonstrated that learning models can achieve accuracy levels close to those of fully-supervised training methods, albeit with only a fraction of the data being annotated. 

In its most basic form, data reduction can be achieved through a random selection of a subset of the data to be exclusively annotated. Random selection, while straightforward and simple to implement, might not capture the most informative data points. Hence, other strategies aim to identify crucial Regions of Interest (ROIs) and patches, optimizing attributes such as representativeness, consistency, typicality, or areas depicting challenging or straightforward phenomena. There is a noticeable intersection between data reduction and MIL techniques, particularly with Expectation Maximization (EM) methods that identify discriminative instances. The key distinction, however, is that data reduction strategies are designed to selectively annotate important instances, whereas MIL approaches are geared towards extracting refined features from these important instances.


\subsubsection{Existing Works} In the broader medical field, there are notable examples of efforts to implement data reduction, such as the works of \citet{39,38}. \citet{38} reduce annotation by 99.9\% for the analysis of Acute Myeloid Leukemia in cytology images by proposing a three-stage process that filters out non-informative regions in WSIs. However, this method may face challenges in handling cluttered nuclei, which are commonly encountered in histopathology WSIs. \citet{39} base the selection on the typicality of the samples, determined by the number of required IHC stains; fewer stains are indicative of more typical samples, while a greater number of stains point to atypical samples. Such criteria for determining typicality might not be universally applicable across all histopathology datasets. It is important to note that this survey primarily focuses on DL tools aimed at facilitating data reduction, whereas the methodologies mentioned above are considered handcrafted.

\begin{figure}[]
    \centering
    \includegraphics[]{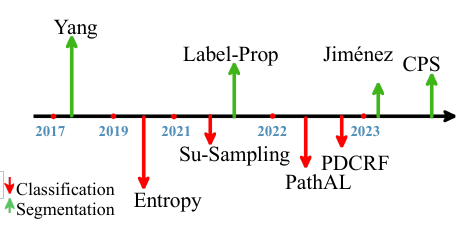}
    \caption{Chronological Trend Line of the Most Relevant Data Reduction Techniques.The length of an arrow is proportional to its popularity per its category.}
    \label{fig:reduction_trend}
\end{figure}

\begin{table*}[ht]
\centering
\caption{Summary of Data Reduction Methods in Histopathology, organized chronologically in descending order. Performance and annotation percentages are calculated w.r.t. the baseline mentioned in the work. Repeated datasets are in boldface. }
\label{tab:bench-reduction}
\begin{tabular}{>{\footnotesize}c>{\footnotesize}c>{\footnotesize}c>{\footnotesize}c>{\footnotesize}c}
\toprule
\textbf{Work} & \textbf{Task} & \textbf{Dataset - Organ} & \textbf{Annot. \%} & \textbf{Perf. \%} \\ \midrule
\multirow{3}{*}{CPS - \citep{44}} & \multirow{3}{*}{SEG} & TCGA - multi-organ & \multirow{2}{*}{5\%} & \textcolor{red}{$\downarrow$}0.34\% (AJI) \textcolor{red}{$\downarrow$}0.28\% (Dice) \\
 &  & TNBC - Breast &  & \textcolor{green}{$\uparrow$}1.7\% (AJI) \textcolor{green}{$\uparrow$}1.5\% (Dice) \\
 &  & MoNu - multi-organ & 7\% & \textcolor{red}{$\downarrow$}1.77\% (AJI) \textcolor{red}{$\downarrow$}1.02\% (Dice) \\ \addlinespace
Jiménez - \citep{176} & SEG & \textbf{Glas - Colorectal ADC} & 26.28\% & \textcolor{red}{$\downarrow$}10.73\% (Dice) \\ \addlinespace
\multirow{3}{*}{PDCRF - \citep{115}} & \multirow{3}{*}{CLS} & \textbf{TCGA - STAD - Gastric} & \multirow{3}{*}{20\%} & \textcolor{red}{$\downarrow$}23.56\% (ACC) \\
 &  & \textbf{TCGA - COAD - Colon} &  & \textcolor{red}{$\downarrow$}24\% (ACC) \\
 &  & \textbf{TCGA - READ - Rectum} &  & \textcolor{red}{$\downarrow$}20.83\% (ACC) \\ \addlinespace
PathAL - \citep{178} & CLS & PANDA Challenge  - Prostate & 28.5\% & \textcolor{red}{$\downarrow$}4.71\% (k) \\ \addlinespace
\multirow{2}{*}{Su-Sampling - \citep{42}} & \multirow{2}{*}{CLS} & \multirow{2}{*}{\textbf{TCGA - avg(STAD,CAOD,READ)}} & \multirow{2}{*}{31\%} & \textcolor{red}{$\downarrow$}18.39\% (ACC) \\
 &  &  &  & \textcolor{red}{$\downarrow$}20\% (ACC) \\ \addlinespace
Label-Prop - \citep{143} & SEG & \textbf{Glas  - Colorectal} & 20\% & \textcolor{red}{$\downarrow$}12.43\% (D) \\ \addlinespace
Entropy - \citep{177} & CLS & BreaKHis - Breast & 20\% & \textcolor{red}{$\downarrow$}8.43\% (ACC) \\ \addlinespace
Yang - \citep{173} & SEG & MICCAI15 - avg(A,B) & 20\% & \textcolor{red}{$\downarrow$}1.47\% (D) \\
\bottomrule
\end{tabular}
\end{table*} 

\citet{173} base the selection on uncertainty, to identify valuable instances, and representativeness, to enhance the generalization and quality of the features extracted. Initially, several Fully Convolutional Networks (FCNs) are trained on a small amount of annotated data. Bootstrapping is then employed to estimate the uncertainty of annotating unlabeled samples, and the top $K$ uncertain samples are chosen to form a subset $S_c$. Subsequently, the most representative subset $S_a \in S_c$ is selected. Specifically, instances from $S_c$ are incrementally added to $S_a$, where each addition is based on evaluating the maximum similarity between $S_a$ members and the remaining unlabeled images, thereby identifying a subset that is both highly uncertain and representative of the whole dataset. Using only 20\% of the annotation budget allocated to a random selection process, this approach reportedly maintains about 99\% of the Dice metric performance for gland segmentation tasks. \citet{177} proposed selection by an entropy-based measurement and a confidence-boosting strategy. Beginning with a CNN trained on minimal data, predictions of high entropy — indicating significant uncertainty in the CNN's prediction and thus signifying a potentially valuable addition to the training set — are marked for future manual annotation, aligning with the strategy of~\citet{173}. In the confidence-boosting approach, predictions made with high confidence by the model are incorporated into the training set with their predicted labels for subsequent stages. Training with only 20\% of data chosen through the entropy-based method can preserve approximately 92\% accuracy, significantly lowering annotation costs.

\citet{143} approached the joint classification and segmentation of histopathology images with their Label-prop method. They aimed to make use of data remaining unlabeled in the AL process in addition to annotating selected candidate samples at each phase. Candidate samples are those close in proximity (based on color distribution) to a fully-annotated image of the same class, rather than relying on confidence levels as per~\citet{173}. The remaining unlabeled images are relayed to expert annotation via random selection. By using only 20\% of the labeled data, they were able to retain around 88\% of the Dice metric performance compared to a fully-supervised model trained with comprehensive pixel-level annotations. Shen and Ke developed a selection mechanism called SU-sampling~\citep{42}, which leverages spatial information and uncertainty to approximate the informativeness of unlabeled data. This method uses two loss terms to measure uncertainty: one is a model-specific uncertainty-like loss applied to unlabeled patches, and the other comes from the main classifier's loss on iteratively updated labeled patches. High loss values indicate regions of uncertainty, prompting the system to consider neighboring unlabeled patches for annotation. To avoid redundancy and ensure a diversified querying across the dataset, SU-sampling includes a counter-measure to decrease the likelihood of selecting patches near previously queried areas. Using approximately 30\% of the dataset, this approach can maintain a high level of accuracy, as shown in Table.~\ref{tab:bench-reduction}.
Li and colleagues introduced an Active Learning (AL) framework tailored for histopathology, named PathAL~\citep{178}. PathAL selects uncertain but informative samples to experts for annotation based on the prediction confidence. The framework differentiates between noisy (mislabelled) samples and hard samples, proposing the integration of only the latter into the training pool due to the detrimental effect of the former. Inspired by Curriculum Classification~\citep{currCls} and using O2U-Net~\citep{o2u}, PathAL ranks the complexity of unlabeled samples and identifies noisy or hard samples by monitoring loss variation, sample complexity, and predictive entropy across training epochs. By annotating informative samples and using confident samples, while excluding noisy ones, PathAL reportedly achieves 100\% performance retention using only 60\% of annotations. \citet{115} proposed the Pathology Deformable Conditional Random Field (PDCRF), which enhances the standard CRF model by integrating a deformation mechanism. This allows the PDCRF to adjust spatial relationships among patches on a WSI, optimizing the selection process based on morphological representativeness and diagnostic relevance. The PDCRF effectively identifies patches that either share high morphological similarities with their vicinity or represent significant pathological changes. With only 20\% of the data used, the method fully preserves the upper-end accuracy as evidenced in Table.~\ref{tab:bench-reduction}.

In their study, \citet{44} introduce the Consistency-based Patch Selection (CPS) strategy, which involves sample selection and data synthesis. The selection is based on representativeness and intra-consistency. The process assesses representativeness via a density-centered approach, identifying samples that are proximal to the centroids of clusters formed within the feature space through k-means clustering. This initial coarse-level clustering is followed by a finer clustering where patches are subdivided into various regions through k-means, to identify consistent patches, as approximated through the clustering's compactness. Intra-consistent patches greatly simplify the training of the subsequent synthesis stage, and the framework demonstrates the potential to match or even exceed the performance of models trained under full supervision. \citet{176} assess the intersection of semi-supervised learning (detailed in Section~\ref{section:semi}) with AL. Their exploration covers three pseudo-labeling techniques alongside four distinct AL strategies including random selection, selection based on the minimal entropy in class distribution, selection based on the lowest prediction confidence, and selection according to the greatest prediction inconsistency when subject to various image disturbances. An initial annotation of about 15\% of what would be required for full pixel-level annotation is used. This amount is then gradually increased within the AL framework. The experiments demonstrated that about 96\% of the Dice score can be achieved while only utilizing about 25\% of the annotation content, as detailed in Table.~\ref{tab:bench-reduction}.

\subsubsection{Summary}
Table.~\ref{tab:bench-reduction} quantitatively assesses various data reduction techniques, with their chronological developments illustrated in Fig.~\ref{fig:reduction_trend}. The essential features of these data reduction approaches are outlined below:

\begin{itemize}
    \item 
 
    At their core, data reduction methods are designed to identify and select critical samples within histopathology datasets for subsequent expert annotation.

    \item 
    According to multiple studies on data reduction, significant savings in annotation costs can be achieved with minimal impact on the performance of DL applications in histopathology. Certain approaches demonstrate less than 1\% drop in performance while being trained on merely 5\% of the total data.

    \item 
    The benefits of employing data reduction techniques are dual: They not only reduce the workload on experts by minimizing the amount of data they need to label but also ensure that the effort put into annotation results in high-quality data, thereby avoiding the issue of Coarse Label (CL) scarcity.

    \item 

    These methods potentially enhance collaboration between DL researchers and pathologists, for instance, by aiding in the creation of expansive histopathology datasets. Here, the annotation priorities set by pathologists could be guided by insights from DL specialists.

    \item 

    Despite their advantages, data reduction strategies have not been as extensively explored or utilized as other methods aimed at overcoming data scarcity, suggesting a fertile area for future research within this domain.

\end{itemize}

\begin{figure*}[t]
\begin{subfigure}{0.5\textwidth}
    \centering
    \includegraphics[width=\textwidth]{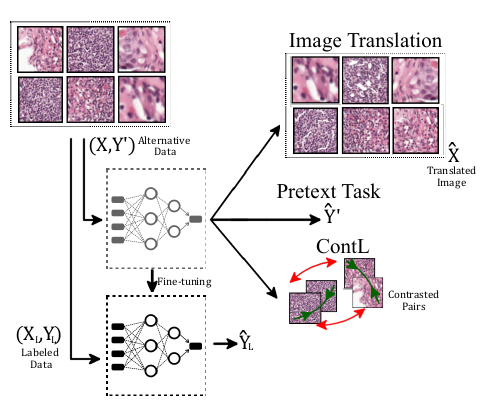}
    \caption{}
    \label{fig:semi-a}
\end{subfigure}
~
\begin{subfigure}{0.5\textwidth}
    \centering
    \includegraphics[width=\textwidth]{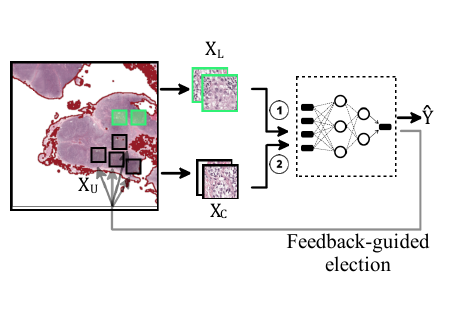}
    \caption{}
    \label{fig:semi-b}
\end{subfigure}
  \caption{The Semi-supervised Learning Paradigm. (a) Few images $X_L$ have their labels $Y_L$ accessible. All images $X$ have proxy labels $Y'$. Images with their proxy labels are inputted to a learner. Image reconstruction produces $\hat{X}$, a pretext task produces $\hat{Y'}$, and contrastive learning pulls/pushes positive/negative pairs. Fine-tuning utilizes the small-scale labeled data $Y_L$ to produce $\hat{Y}$. (b) An illustration of the pseudo labeling approach. The learner is trained first on the small-scale labeled data $X_L$, which elects candidate samples $X_C$ from the unlabeled set $X_U$, and predicts their pseudo-label.}
  \label{fig:semi}  
\end{figure*}


\section{Data Altering (DA) Methods}
In this section, we explore Data Altering (DA) strategies, crucial for augmenting the volume and variety of training data available for DL models under conditions of data scarcity. We focus on three primary categories within this approach: Semi-supervised Learning, Data Augmentation/Synthesis, and Transfer Learning. Each category illustrates how these strategies can effectively expand the dataset's scope, enhancing model robustness and reducing the overfitting risks associated with small datasets.

\subsection{Semi-supervised Learning}
\label{section:semi}
In semi-supervised learning, both labeled ($X_L$) and unlabeled data ($X_U$) are utilized to form the training pool $X = X_L \cup X_U$. This approach is particularly valuable in scenarios where data is plentiful but annotation is prohibitively costly or challenging to secure, thereby causing the volume of $X_U$ to far exceed $X_L$. This methodology hinges on techniques that empower models to derive insights and develop capabilities from the vast reserves of unlabeled data. In this review, we have opted not to allocate a specific section to unsupervised methods, stemming from the assumption that a minimal set of labeled data is generally available, which allows for subsequent refinement post the unsupervised phase, thereby aligning more closely with semi-supervised paradigms.


\subsubsection{Self-supervision}
In self-supervised learning, a learner $\mathscr{F}$ is trained on the training set $X$ without accessing its labels $Y$. This training is typically conducted in one of three primary ways. The first involves creating a set of proxy labels $Y'$, characterized by their abundance, accessibility, automatic generation, or derivation directly from $X$. Consequently, the learner undertakes the surrogate task $X \xrightarrow{\mathscr{F}} \hat{Y'}$, a process known as self-supervision on pretext tasks. By predicting labels for relevant tasks using the same input $X$ and leveraging the breadth provided by extensive unlabeled data training, the model learns representations of $X$ that are beneficial for subsequent histopathology tasks by allowing for weight fine-tuning using a small labeled dataset $(X_L, Y_L)$, thus reducing overfitting while obviating the need for costly expert annotations. Notably, common \textit{pretext tasks} such as unrotating an image~\citep{unrotate} or reassembling jigsaw puzzles~\citep{jigsaw} from image tiles are not expected to translate well to histopathology due to the inherently complex orderings and rotations in histopathology images, which lack a standard global orientation. 

The second approach to self-supervision involves an image translation task $X \xrightarrow{\mathscr{F}} \hat{X}$, where the model is pre-trained. This translation, which can include image reconstruction, generation, or modification, aims to foster meaningful representations of $X$. This contrasts with data augmentation strategies (elaborated in Section. \ref{sec:aug}), where synthetic images $\hat{X}$ are added to the training pool, i.e., $X \leftarrow X \cup \hat{X}$.

Lastly, Contrastive Learning (ContL) is a widely adopted self-supervised technique in which the model learns to classify embeddings of input samples into positive and negative pairs. Specifically, the learner's backbone is trained to project images $X \xrightarrow{\mathscr{F}} Z$, where $Z$ resides in a feature space designed to bring positive pairs closer and keep negative pairs apart. These self-supervised methods are depicted in Fig. \ref{fig:semi-a}. It is worth noting that it is typical to consider image translation and ContL methods as pretext tasks, but we choose to separate them in this analysis for clarity.

\paragraph{\textbf{Histopathology Specific Pretext Tasks}}
\citet{129} propose the count of nuclei as the supervisory signal for a pretext task, preceding the original task of detection through classification. The backbone learns to rank pairs of patches into two classes: consisting of more nuclei vs. consisting of less nuclei. The authors suggest to construct each pair by cropping a small area out of existing image patches, thus rendering a patch with a count of nuclei less than or equal to the original one. Around 91\% of classification accuracy is retained when compared to a fully supervised model (see Table. \ref{tab:semi}), despite only finetuning the weights on 2 labeled images. Koohbanani and colleagues~\citep{85} proposed Self-Path which incorporates both, general self-supervision tasks originally proposed for natural images, and novel pretext tasks specific to histopathology. The general pretext tasks include the prediction of image rotation and flipping, image reconstruction, adversarial image-synthesis with real/fake prediction, and source vs. target domain prediction. The pathology-specific tasks include magnification prediction, the so-called JigMag which mimics the jigsaw puzzle task~\citep{jigsaw} and the prediction of the Hematoxylin channel of the WSI. Experiments document that the performance of a fully-supervised baseline model is almost entirely retained by integrating the proposed unsupervised pretext tasks, and only fine-tuning the weights on 20\% of the labeled data. 

\begin{figure*}[]
    \centering
    \includegraphics[width=\linewidth]{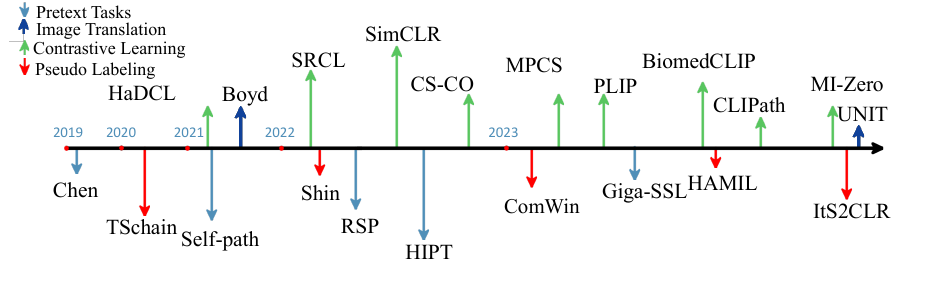}
    \caption{Chronological Trend Line of the Most Relevant Semi-supervised Methods. The length of an arrow is proportional to its popularity per its category. All categories are frequently used, with ContL methods being the most popular.}
    \label{fig:semi_trend}
\end{figure*}
Srinidhi and peers propose the Resolution Sequence Prediction (RSP) as a surrogate task~\citep{rsp}. Specifically, a tuple of three patches representing the same region at different magnifications is inputted to a Siamese CNN trained to order them in terms of magnification. A teacher module pseudo-labels unlabeled samples which a student model learns to match the predictions of. A consistency loss is also employed to enforce matching the prediction of images with their augmented counterparts.

The recent works of \citet{200} and \citet{156} have the particularity of applying self-supervision at the level of the slide. The Hierarchical Image Pyramid Transformer (HIPT)~\citep{200} is composed of a pyramid of three stacked vision transformers (ViTs), analyzing over different magnifications (slide, region, and patch levels), and trained following the student-teacher Self-Distillation with No Labels (DINO)~\citep{dino} paradigm. The student, which only accesses local crops, learns to imitate the teacher which has access to global crops allowing it to make better classification predictions, thus concentrating the attention in the Transformer model on the class-bearing regions. In the slide-level classification task, the HIPT model outperforms the baseline which is trained in a fully-supervised fashion, despite only fine-tuning the weights on 25\% of the data. 

In the Giga-SSL~\citep{156} model, a WSI $X$ is divided into $T$ tiles, where slide-level and tile-level color/geometric perturbations are applied. The used SparseConvMIL~\citep{sparseconvmil} method extracts features from images and confer them with spatial information; specifically x-y coordinates to build a 2D map of latent features. Said features are combined into a global representation using an aggregation module. MoCo~\citep{moco}, a ContL framework, is used to pre-train the encoder module. Giga-SSL demonstrates a greater performance improvement than HIPT as evidenced in Table.\ref{tab:semi}.

\paragraph{\textbf{Image Translation}} 

\citet{124} input crops of patches to a generative network, and train it on the pretext task of outpainting, which can be considered a form of masked image translation. The model learns meaningful representations that enable it to synthesize realistic and detailed expansions, featuring different histopathological objects and tissue details.

It is noted that the majority of the reviewed work tackles classification. In contrast, \citet{154} tackle the segmentation of nuclei in histopathology images with their Unpaired nucleus-aware Image-to-image Translation (UNIT) task. They use a CycleGAN~\citep{cyclegan} to transform randomly generated polygons into histopathology images containing nuclei, which are reused with a segmentation pipeline to pre-train the weights involved in the segmentation network. As such, the encoder and decoder of the generator, as well as the segmentation branch, are all pre-trained in the UNIT framework, and can be fine-tuned on the small-scale annotated data.



\paragraph{\textbf{Contrastive Learning}}

The literature contains diverse examples aimed at devising a version of the existing data that observes a positive/negative pairing which can then be used to learn a contrastive task. Popular ContL frameworks such as SimCLR~\citep{simclr} and MoCo~\citep{moco} have been applied to histopathology images. Ciga and colleagues follow the SimCLR framework \citep{195}. They augment the original image with several perturbations, whereby the off-springs are positively paired with the image in the feature domain, while the off-springs across different source images are negatively paired and are conversely pushed apart. By only using 20\% of the labeled data for fine-tuning, a classifier pre-trained by the proposed ContL framework may outperform the same classifier trained on the full dataset, only pre-trained on Imagenet (see Table. \ref{tab:semi}). \citet{196} experiment with Moco~\citep{moco} and RSP~\citep{rsp} and integrate a curriculum learning approach in the fine-tuning stage. Consequently, the proposed framework, dubbed Hardness-aware Dynamic Contrastive Learning (HaDCL), outperforms the baseline that is randomly initialized and trained on the same amount of data. 

In contrast, the pairs according to \citet{134} is defined via a cosine similarity index, rendering positive pairs from within the same WSI as well as accross different WSIs. The proposed Semantically Relevant Contrastive Learning (SRCL) method, is evaluated across five different histopathology tasks and demonstrate great robustness and transferability compared to other compared self-supervised learning techniques. \citet{128} porpose the Magnification Prior Contrastive Similarity (MPCS), which positively pairs two images of the same view at different magnification factors. In comparison, \citet{63} propose CS-CO, where stains of images are separated into Hematoxylin and Eosin channels, and two autoencoders are trained to construct the E-channel image from the H-channel, and vice-versa. Afterwards, a ContL stage adopts a Siamese NN, whereby the encoders extract features from the stain-separated images, concatenate and project them into a latent space in which augmented counter-parts are pulled closer.


\paragraph{\textbf{Vision-Language Contrastive Learning}}
Recent advances pair across image-text domains, which represents a cost-effective alternative to expert annotation, especially as modern automatic tools substantially simplify the text mining process. The OpenPath~\citep{plip} dataset includes approximately 200,000 histopathological image-text pairs, the ARCH dataset contains about 15,000 pairs~\citep{180}, and the Quilt-1M dataset boasts 1 million pairs~\citep{quilt}. The pioneering work in Contrastive Language-Image Pre-training (CLIP) set the stage for zero-shot transfer from text to images~\citep{clip}. Following this framework, \citet{plip} introduced Pathology Language–Image Pre-training (PLIP) specifically for histopathology images, leveraging textual information mined from medical-related tweets and their responses to compile a dataset of image-text pairs. The process integrates a contrastive loss which trains image and text encoders to synchronize image embeddings with corresponding text embeddings. Subsequently, textual class prompts, such as ``an image showing metastasis," are processed by the text encoder and compared via cosine similarity to the embeddings of a test image produced by the image encoder. Experiments highlight the robustness of this framework, where the highest cosine similarity consistently corresponds with the prompt denoting the correct class of the image.

\citet{biomedclip} developed the BiomedCLIP model, which classifies patch-level histopathology images using text captions sourced from biomedical research articles in PubMed Central. In a similar vein, the CLIPPath model proposed by \citet{175} freeze the CLIP module and attach to it convolutional layers with skip connections that are fine-tuned on histopathology images. \citet{151} extend the CLIP framework to the MIL setting with their MI-Zero model, which performs slide-level classification by aggregating zero-shot patch-level predictions using various pooling mechanisms. Table. \ref{tab:semi} summarizes the performance enhancements of these methods across various datasets, all benchmarked against the foundational CLIP model. 

\begin{table*}[]
\centering
\caption{Summary of Relevant Semi-supervised Learning Approaches by Type, organized chronologically in descending order. (-) indicates unreported value. CL refers to Coarse Labels.}
\label{tab:semi}
\begin{tabular}{>{\footnotesize}c>{\footnotesize}c>{\footnotesize}c>{\footnotesize}c>{\footnotesize}c>{\footnotesize}c}
\textbf{Type} & \textbf{Work} & \textbf{Task} & \textbf{Dataset} & \textbf{Annot. \%} & \textbf{Perf. \%} \\ \hline
\multirow{10}{*}{\rotatebox[origin=c]{30}{Pretext Tasks}} & \multirow{3}{*}{Giga-SSL - \citep{156}} & \multirow{3}{*}{CLS} & \textbf{TCGA - Breast} & \multirow{3}{*}{25\%} & \textcolor{green}{$\uparrow$}14.4\% (AUC) \\
 &  &  & \textbf{TCGA - Lung} &  & \textcolor{green}{$\uparrow$}5.27\% (AUC) \\
 &  &  & \textbf{TCGA - Kidney} &  & \textcolor{green}{$\uparrow$}1.67\% (AUC) \\ \addlinespace 
 & \multirow{3}{*}{HIPT - \citep{200}} & \multirow{3}{*}{CLS} & \textbf{TCGA - Breast} & \multirow{3}{*}{25\%} & \textcolor{green}{$\uparrow$}5.53\% (AUC) \\
 &  &  & \textbf{TCGA - Lung} &  & \textcolor{green}{$\uparrow$}3.48\% (AUC) \\
 &  &  & \textbf{TCGA - Kidney} &  & \textcolor{green}{$\uparrow$}1.56\% (AUC) \\ \addlinespace 
 & \multirow{2}{*}{RSP - \citep{rsp}} & \multirow{2}{*}{CLS} & \textbf{Kather - Colon} & \multirow{2}{*}{25\%} & \textcolor{red}{$\downarrow$}0.1\% (AUC) \textcolor{red}{$\downarrow$}0.76\% (ACC) \\
 &  &  & \textbf{Camelyon16 - Breast} &  & \textcolor{green}{$\uparrow$}6.01\% (AUC) \textcolor{red}{$\downarrow$}3.41\% (ACC) \\ \addlinespace 
 & Self-path - \citep{85} & CLS & \textbf{Camelyon16 - Breast} & 20\% & \textcolor{red}{$\downarrow$}0.54\% (AUC) \\ \addlinespace 
 & Chen - \citep{129} & CLS & \textbf{Camelyon16 - Breast} & 30\% & \textcolor{red}{$\downarrow$}4.22\% (ACC) \\ \hline
\multicolumn{1}{>{\footnotesize}l}{\multirow{4}{*}{\rotatebox[origin=c]{30}{Image Translation}}} & \multirow{2}{*}{UNIT - \citep{154}} & \multirow{2}{*}{SEG} & Lizard - DPath & \multirow{2}{*}{25\%} & \textcolor{red}{$\downarrow$}15.76\% (D) \\
\multicolumn{1}{l}{} &  &  & Lizard - CRAG &  & \textcolor{red}{$\downarrow$}35.32\% (D) \\ \addlinespace 
\multicolumn{1}{l}{} & \multirow{2}{*}{Boyd - \citep{124}} & \multirow{2}{*}{CLS} & \textbf{Camelyon17 - Breast} & \multirow{2}{*}{-} & \textcolor{red}{$\downarrow$}5.71\% (ACC) \textcolor{red}{$\downarrow$}6.14\% (F) \\
\multicolumn{1}{l}{} &  &  & \textbf{NCT-CRC-HE - Colorectal} &  & \textcolor{red}{$\downarrow$}1.33\% (ACC) \textcolor{red}{$\downarrow$}1.11\% (F) \\ \hline
\multirow{25}{*}{\rotatebox[origin=c]{30}{Contrastive Learning}} & \multirow{3}{*}{MI-Zero - \citep{151}} & \multirow{3}{*}{CLS} & Private - Breast & \multirow{3}{*}{Zero-shot} & \textcolor{green}{$\uparrow$}34.4\% (ACC) \\
 &  &  & Private - Lung &  & \textcolor{green}{$\uparrow$}40\% (ACC) \\
 &  &  & Private - Kidney &  & \textcolor{green}{$\uparrow$}120.1\% (ACC) \\ \addlinespace 
 & \multirow{2}{*}{CLIPath - \citep{175}} & \multirow{2}{*}{CLS} & \textbf{MHIST} & 1\% & \textcolor{green}{$\uparrow$}73.2\% (ACC) \textcolor{green}{$\uparrow$}28.3\% (AUC) \\
 &  &  & \textbf{PCam} & 0.1\% & \textcolor{green}{$\uparrow$}35.2\% (ACC) \textcolor{green}{$\uparrow$}41.5\% (AUC) \\ \addlinespace 
 & \multirow{4}{*}{BiomedCLIP - \citep{biomedclip}} & \multirow{4}{*}{CLS} & \textbf{PCam} & \multirow{4}{*}{Zero-shot} & \textcolor{green}{$\uparrow$}15.1\% (ACC) \\
 &  &  & LC25000 - Lung &  & \textcolor{green}{$\uparrow$}112.4\% (ACC) \\
 &  &  & LC25000 - Colon &  & \textcolor{green}{$\uparrow$}24.8\% (ACC) \\
 &  &  & TCGA - TIL &  & \textcolor{green}{$\uparrow$}24.8\% (AUC) \\ \addlinespace 
 & \multirow{4}{*}{PLIP - \citep{plip}} & \multirow{4}{*}{CLS} & \textbf{Kather - Colon} & \multirow{4}{*}{Zero-shot} & \textcolor{green}{$\uparrow$}318.5\% (F) \\
 &  &  & PanNuke &  & \textcolor{green}{$\uparrow$}86.36\% (F) \\
 &  &  & DigestPath &  & \textcolor{green}{$\uparrow$}2673\% (F) \\
 &  &  & \textbf{WSSS4LUAD} &  & \textcolor{green}{$\uparrow$}52.5\% (F) \\ \addlinespace 
 & \multirow{2}{*}{MPCS - \citep{128}} & \multirow{2}{*}{CLS} & BreaKHis - Breast & \multirow{2}{*}{20\%} & \textcolor{red}{$\downarrow$}4.95\% (ACC) \\
 &  &  & \textbf{BACH} &  & \textcolor{red}{$\downarrow$}21.55\% (ACC) \\ \addlinespace 
 & CS-CO - \citep{63} & CLS & \textbf{NCT-CRC-HE - Colorectal} & 10\% & \textcolor{green}{$\uparrow$}2.24\% (ACC) \\ \addlinespace 
 & \multirow{3}{*}{SimCLR - \citep{195}} & \multirow{3}{*}{CLS} & \textbf{BACH} & \multirow{3}{*}{20\%} & \textcolor{red}{$\downarrow$}10.96\% (F) \\
 &  &  & \textbf{NCT-CRC-HE} &  & \textcolor{green}{$\uparrow$}10\% (F) \\
 &  &  & Lymph Dataset &  & \textcolor{green}{$\uparrow$}8.33\% (F) \\ \addlinespace 
 & \multirow{4}{*}{SRCL - \citep{134}} & \multirow{4}{*}{CLS} & \textbf{NCT-CRC-HE} & 10\% & \textcolor{green}{$\uparrow$}1.87\% (ACC) \textcolor{green}{$\uparrow$}1.34\% (F) \\
 &  &  & \textbf{Camelyon16 - Breast} & \multirow{3}{*}{CL} & \textcolor{green}{$\uparrow$}10.16\% (ACC) \textcolor{green}{$\uparrow$}7.9\% (AUC) \\
 &  &  & \textbf{TCGA - Lung} &  & \textcolor{green}{$\uparrow$}6.17\% (ACC) \textcolor{green}{$\uparrow$}3.73\% (AUC) \\
 &  &  & \textbf{TCGA - Kidney} &  & \textcolor{green}{$\uparrow$}4.99\% (ACC) \textcolor{green}{$\uparrow$}0.41\% (AUC) \\ \addlinespace 
 & \multirow{2}{*}{HaDCL - \citep{196}} & \multirow{2}{*}{CLS} & \textbf{Camelyon16 - Breast} & \multirow{2}{*}{100\%} & \textcolor{green}{$\uparrow$}17.23\% (ACC) \textcolor{green}{$\uparrow$}20.77\% (AUC) \\
 &  &  & \textbf{MHIST} &  & \textcolor{green}{$\uparrow$}2.74\% (ACC) \textcolor{green}{$\uparrow$}1.82\% (AUC) \\ \hline
\multirow{6}{*}{\rotatebox[origin=c]{30}{Pseudo Labeling}} & ItS2CLR - \citep{204} & CLS & \textbf{Camelyon16 - Breast} & CL & \textcolor{red}{$\downarrow$}1.25\% (AUC) \textcolor{red}{$\downarrow$}0.61\% (F) \\ \addlinespace 
 & \multirow{2}{*}{HAMIL - \citep{203}} & \multirow{2}{*}{SEG} & LUAD-HistoSeg & \multirow{2}{*}{CL} & \textcolor{green}{$\uparrow$}17.57\% (D) \\
 &  &  & \textbf{WSSS4LUAD (subset)} &  & \textcolor{green}{$\uparrow$}17.46\% (D) \\ \addlinespace 
 & ComWin - \citep{206} & SEG & Colon & 5\% & \textcolor{red}{$\downarrow$}26.95\% (D) \textcolor{red}{$\downarrow$}32.49\% (J) \\ \addlinespace 
 & Shin - \citep{133} & CLS & \textbf{Camelyon16 - Breast} & 20\% & \textcolor{red}{$\downarrow$}1.8\% (AUC) \\ \addlinespace 
 & TSchain - \citep{TSchain} & CLS & \textbf{NCT-CRC-HE} & 20\% & \textcolor{green}{$\uparrow$}0.48\% (ACC)
\end{tabular}
\end{table*}


\subsubsection{Pseudo Labeling}
\label{section:semi-pseudo}

Pseudo labeling methods are designed to iteratively refine tentative labels generated by weakly-trained learners, as shown in Fig. \ref{fig:semi-b}. Specifically, the process starts with a learner $\mathscr{F}: X \rightarrow \hat{Y}$, using initially the training dataset $(X,Y) = (X_L,Y_L)$, where $(X_L,Y_L)$ represent the small set of labeled images and their corresponding labels. As training progresses, pseudo labeling techniques utilize the model's acquired knowledge to infer labels for the unlabeled set, thereby expanding $(X,Y)$ to include $(X,Y) = (X_L,Y_L) \cup (X_c,\hat{Y_C})$, with $X_c$ being candidate instances from the unlabeled data pool $X_U$. A criterion, such as high-confidence predictions, guides the inclusion of these labels into the training set in a repetitive, potentially refining manner until training stabilizes.

In the context of histopathology images, pseudo labeling proves beneficial even when addressing the scarcity of the second type of weak annotation, namely Coarse Labeling (CL), as detailed in Section.\ref{section:intro-digital}. A learner trained on coarse labels is capable of generating fine-level pseudo-labels, epitomizing a weak learner due to the misalignment between the coarseness of the data and the desired detailed granularity. It is important to note the similarities between pseudo labeling and Active Learning (AL) methods discussed in Section.\ref{sec:reduction}. The primary distinction lies in the source of new labels; active learning relies on expert feedback, whereas pseudo labeling depends on the incrementally improving model. In this vein, Belharbi \textit{et al.}~\citep{143} devise an AL framework that incorporates a pseudo labeling stage, enabling the model to annotate samples close to those already labeled.

Moreover, the concept of co-training~\citep{cotraining} expands on these ideas by training multiple models on a diversified labeled dataset, where each model assists in annotating the unlabeled set for the others. This diversity, which can arise from using different stains (e.g., H\&E vs. IHC or fluorescence in situ hybridization) or magnification levels, helps to mitigate the impact of noisy labels as the noise in one view does not necessarily carry over through successive training iterations. Similarly, tri-training~\citep{tritraining}, also known as co-regularization, employs three or more learners and penalizes discrepancies in the pseudo annotations produced by any two of them. This method tends to be more resilient to the adverse effects of noisy or incorrect predictions by any single model involved.

\citet{135} assign pseudo-labels based on the confidence of predictions made by the learner. Namely, if the probability $P$ of predicting a sample as positive is $\geq 0.9$, it is assigned a pseudo-label of 1. On the other hand, $P \leq 0.1$, it is assigned a pseudo-label of 0. This iterative process is carried out several times until convergence, effectively increasing the confidence of the model's predictions as the training continues. \citet{TSchain} propose the Teacher-Student Chain (TS-Chain), where the teacher is trained on the small-scale data $X_l, |X_l| = N$, and $Y_l$ are the labels associated with $X_l$ and produces $Y_u$; the labels of the rest of the large cohort of unlabeled images $X_u, |X_u| = M, N << M$. The student is trained on $X_u$ and fine-tuned on $X_l$, repeating this dynamic for multiple iteration, where a student becomes the teacher of a new student at each stage.

A prominent problem in the domain of pseudo labeling is erroneous first predictions of the weak learner, which are susceptible to be propagated to later stages. To that end, \citet{133} propose a refinement procedure based on graph-segmentation, where the WSI is treated as a graph with pseudo-labels being its segmentation seeds. Then, a graph-cut algorithm is applied to refine the pseudo-labels and obtain enhanced label suggestions, taking into consideration the confidence of predictions made by the initial model. \citet{207} dilute the effect of potentially erroneous initial pseudo labels by incorporating a KNN-based super-instance formulation through their Learning from Noisy Pseudo Labels (LNPL) method. Following a similar concept, \citet{206} propose Compete to Win (ComWin) which is grounded in the co-training paradigm. The core strategy is the evaluation of confidence maps from $N$ weak annotators, all maintaining an identical structure initialized differently, and selecting the most confident prediction among them. This represented a dynamic thresholding reliant on the overall confidence of the learners, which escalates as training progresses, rather than relying on a fixed, arbitrary threshold value.

\citet{203} used CAM-based localization -- enhanced by multi-layer fusion and Monte-Carlo augmentation -- to obtain pixel-level pseudo labels from image-level annotations. Then, a tri-training-based approach, named High-resolution Activation Maps and Interleaved Learning (HAMIL) employs three models with similar architectures, whereby the agreement of annotation by the other two models indicates the reliability of the pseudo label that will be used to train the third model. \citet{204} utilize pseudo labels to guide ContL, thus addressing the challenge where the majority of instances in a WSI are typically negative and often from the same class. The inherent mechanism of ContL pulls an image closer to its augmented versions while distancing it from other instances, irrespective of their class affiliation with the anchor sample. To address this, the proposed Iterative Self-paced Supervised Contrastive Learning for MIL Representations (ItS2CLR) evaluates the predictions from an MIL decision-level aggregator, which delivers instance-level predictions that are used to generate instance-level pseudo labels. Concretely, a closed feedback loop is established between the MIL aggregator and the ContL-based feature extractor, whereby better features are extracted in the ContL environment due to the improved pseudo labels generated by the MIL aggregator, and vice versa. On the Camelyon16 dataset, training a learner on only WSI-level labels using the proposed framework obtains the same performance as training a learner with access to instane-level labels, indicating a significant reduction in the annotation burden.

\subsubsection{Summary}
The Camleyon16/17~\citep{cam16,cam17} datasets, along with data extracted from the TCGA~\citep{tcga} record, are frequently employed by existing semi-supervised methods, in conjunction with several other datasets. Table.\ref{tab:semi} compares these methods, which are also chronologically illustrated in Fig.~\ref{fig:semi_trend}. Below, key points summarize the essential aspects of semi-supervised techniques in histopathology:

\begin{itemize}
    \item Semi-supervised learning methods generally aim to utilize the informational value of unlabeled data by associating it with alternate labels. These labels are typically generated through pseudo labeling or by setting up a surrogate task where labels are easier to obtain compared to the primary task.

    \item There are similarities between pseudo labeling, data reduction methods, and some MIL approaches, primarily in their objective to identify and utilize key instances. While MIL methods focus on extracting features from these key instances, data reduction methods involve presenting these key instances to an expert for annotation, and pseudo labeling techniques derive labels for these instances using the learning model.

    \item A significant trend in semi-supervised learning is the adoption of the contrastive learning paradigm, particularly in recent advancements that incorporate vision-language models. These models have shown substantial performance improvements and offer a practical solution to the cumbersome task of annotating in the target domain, thus providing a valuable tool for addressing data scarcity.
    
\end{itemize}

\begin{figure*}[]
        \centering
        \begin{subfigure}[b]{0.475\textwidth}
            \centering
            \includegraphics[width=\textwidth]{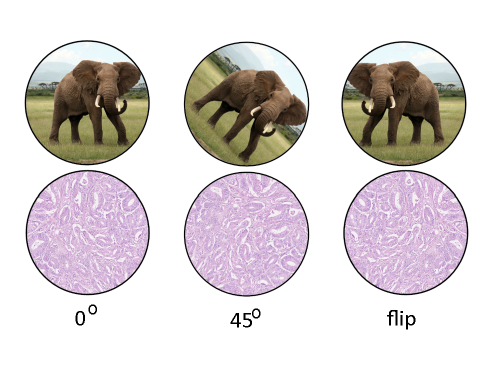}
            \caption[]%
            {{}}    
            \label{fig:aug-class}
        \end{subfigure}
        \hfill
        \begin{subfigure}[b]{0.475\textwidth}  
            \centering 
            \includegraphics[width=\textwidth]{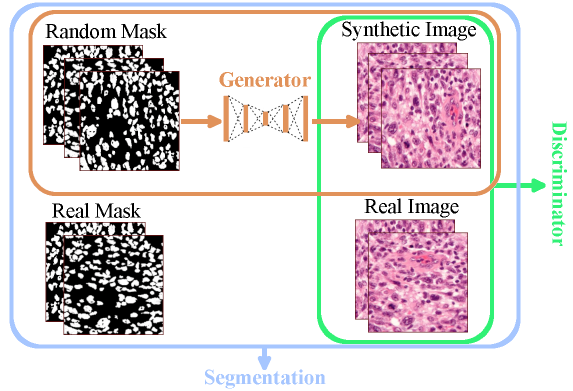}
            \caption[]%
            {{}}    
            \label{fig:aug-gan}
        \end{subfigure}
        \vskip\baselineskip
        \begin{subfigure}[b]{0.475\textwidth}   
            \centering 
            \includegraphics[width=\textwidth]{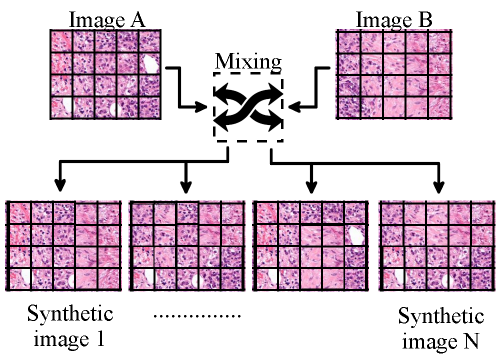}
            \caption[]%
            {{}}    
            \label{fig:aug-mix}
        \end{subfigure}
        \hfill
        \begin{subfigure}[b]{0.475\textwidth}   
            \centering 
            \includegraphics[width=\textwidth]{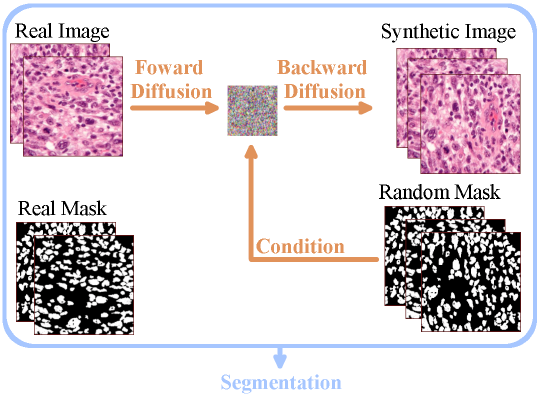}
            \caption[]%
            {{}}    
            \label{fig:aug-diff}
        \end{subfigure}
        \caption[ Simplified Illustration of Data Augmentaiton and Synthesis in Pathology. (a): Classical augmentation techniques applied to histopathology, and natural images. Different orientations diversify the training data in natural images, but not in histopathology. (b): An example of the GAN paradigm, consisting of a generator-discriminator network to synthesize new images from masks, which are added to the real data to learn segmentation. (c): An example of mixing in the image space, which produces N synthetic images from a linear combination of 2 real images. (d): An example of the LDM. The forward diffusion collapses the real image into noise, to which a condition (herein a segmentation mask) is added. A synthetic image that corresponds to the mask and that resembles the original image is produced. Hybrid data consisting of real and synthetic images can be used to learn segmentation.]
        {\small Simplified Illustration of Data Augmentaiton in Pathology. (a): Classical augmentation techniques applied to histopathology, and natural images. Different orientations diversify the training data in natural images, but not in histopathology. (b): An example of the GAN paradigm, consisting of a generator-discriminator network to synthesize new images from masks, which are added to the real data to learn segmentation. (c): An example of mixing in the image space, which produces N synthetic images from a linear combination of 2 real images. (d): An example of the LDM. The forward diffusion collapses the real image into noise, to which a condition (herein a segmentation mask) is added. A synthetic image that corresponds to the mask and that resembles the original image is produced. Hybrid data consisting of real and synthetic images can be used to learn segmentation.}         
    \end{figure*}
    


\subsection{Data Augmentation and Synthesis}
\label{sec:aug}

Data augmentation techniques expand the labeled training data $(X_L,Y_L)$ through computational manipulations. Ideally, a data augmentation pipeline should introduce important phenomena, under-represented by the original data. Consequently, an experienced and more general learner is obtained without incurring additional data collection costs. Data augmentation demonstrates a tremendous performance boost on many tasks pertaining to natural images. Classical data augmentation techniques are well-established, and may include geometric, spatial and color variations, such as warping, rotation, scaling, cropping and merging, or chromatic and stylistic transformations. Such \textit{data-modification} techniques can also be used on histopathology images and have shown some potential. Beyond that, advanced techniques are based on \textit{modern} paradigms that lend themselves useful for histopathology images. Data augmentation can be generally represented by Equation. \ref{eq:aug}.
\begin{equation}
\label{eq:aug}
    X' = \mathscr{F}(X) = t_k \circ t_{k-1} \circ \cdots \circ t_2 \circ t_1(X).
\end{equation} 
The augmentation network $\mathscr{F}$ is a set of transformations applied to each sample in $X$ with a given probability. The output $X'$ is $N'$ augmented data samples, where typically $N' > N$. The mathematical expression includes both subcategories of data augmentation, by interpreting $\mathscr{F}$ as either the composition of several transformations, or the inner operations of a generative model, both resulting in augmented data samples. 


\subsubsection{Considerations for Data Augmentation}

Classical and modern augmentation techniques are not mutually exclusive; rather, many existing approaches integrate both. Nonetheless, it is noted that both categories introduce certain consequences that should be taken into account. Most classical methods are easy to understand, and are integrable in an over-the-shelf fashion, and have the advantage of preserving their annotation information. 

Nevertheless, caution must be practiced when using them as they may introduce some complications when applied to WSIs. Particularly, WSIs inherently lack a global orientation; thus, simple rotation or mirroring may not introduce new, meaningful features but rather perpetuate existing biases and noise. This can undermine the model's generalizability and introduce errors in downstream tasks. For instance, training a classifier on natural images, such as variously oriented images of a sub-Saharan elephant, enhances its generalizability by recognizing these images as belonging to the same class. In stark contrast, due to the absence of a defined global orientation in histopathology images, their rotated versions fail to enrich the classifier’s knowledge base, potentially rendering such augmentations futile or even detrimental, as they might amplify latent noise and biases (see Fig. \ref{fig:aug-class}). Given the complexities of tasks like cancer diagnosis, it is imperative to cultivate a dataset that encompasses a broad spectrum of potential phenomena, underscoring the critical need for diverse data. Although the literature contains examples where classical augmentation enhances the learner's accuracy, more refined techniques often yield superior enhancements in performance. 

Additionally, elastic deformation, as is applied in natural images can greatly affect cellular structures and alter the prognosis associated with the histopathology image. In~\citep{elastic}, elastic deformations improved the glomeruli segmentation; however, subject to meticulous parameter tuning. On the other hand, improperly applied deformations in a gland segmentation dataset, for instance, could severely misrepresent gland morphology, drastically skewing cancer severity assessments. 

Moreover, color or style variation are extensively used in the literature e.g. for color normalization, and have demonstrated great performance boost. However, naive color variation, especially if applied in an inconsistent manner may distort the semantics of histopathology images, and lead to significant diagnostic inaccuracies, particularly in evaluating the extracellular matrix (ECM) components like collagen and fibronectin. In H\&E staining, where the pink to red hues indicate the presence and density of ECM, arbitrary alterations in color tint could misrepresent these critical structures. For example, a lighter or differently tinted pink might suggest a less dense collagen matrix, potentially misleading clinicians into underestimating the severity of the tumor microenvironment. Conversely, contemporary methods that generate high-fidelity synthetic images allow for nuanced customization, at the expense of necessitating sophisticated image selection algorithms, filtering processes, and domain adaptation strategies. Ultimately, the best techniques balance simplicity with precision, as extensively documented in scholarly works. In our analysis, we review classical augmentation methods, in addition to a rich diversity of modern techniques, including Generative Adversarial Networks (GANs), Mixing techniques, and diffusion models.


\subsubsection{Classical Techniques}
Earlier methods in histopathology image synthesis consisted of simple methods which do not usually involve a learnable, rather a deterministic approach. For instance, the early work of \citet{classical15} employs region filling, in which a library of H\&E histopathology textures are used to fill in manually defined regions on an empty canvas. \citet{classical17} augment a metastasis detection dataset through flipping, rotation, chromatic variation, and geometric jitter. \citet{classical18} deconvolve the RGB WSI into the three channels of Hematoxylin, Eosin, and a residual channel, and apply a random affine transformation to each channel individually. Additionally, they mimic the out-of-focus effect by running the images through a Gaussian filter. \citet{60} apply several augmentations towards the efficient classification of prostate histopathology images, including random geometric deformations and PCA-based jitter on the RGB pixels. 

In a relatively recent work by \citet{62}, instead of synthesizing completely new images, the authors propose STRAP (Style TRansfer Augmentation for histoPathology), which transfers style from randomized natural images to existing histopathology images, thus learning domain-agnostic features that generalize better to unseen data undergoing domain shift due to staining elements, imaging protocols...etc (i.e. while avoiding obtaining extra labeled data by an expert). The transfer follows the AdaIN~\citep{adain} framework, in which channel-wise mean and variance are transferred from a style image to a content image, both encoded through an encoder. When testing on unseen data, the method performs favorably to a baseline employing only stain normalization as is shown in Table. \ref{tab:aug}. 

\citet{187}, applied classical augmentation methods such as rotation and colorization; however, guided by an auto-encoder which searches for and ranks the best augmentation policies, instead of randomzing the process. The augmentation takes place at the nucleus level instead of the WSI or patches level, rendering diverse and high-quality synthetic images that improved a detector's performance. Their method is applied to general microscopical data, but it can be useful for H\&E stained WSIs. 


\subsubsection{Generative Adversarial Models}
Generative Adversarial Networks (GANs) \citep{gan} have demonstrated remarkable capabilities in image synthesis. The inherent adversarial training approach of GANs encourages the generator module to create highly realistic images designed to deceive the discriminator module, which is simultaneously trained to differentiate between synthetic and real images. This dynamic is typically represented as a min-max game, as depicted in Eq. \ref{eq:gan} \citep{gan}, and illustrated in Fig. \ref{fig:aug-gan}, where $\mathscr{G}$ symbolizes the generator module, and $\mathscr{D}$ represents the discriminator module. The literature features innovative contributions using GANs, which include generating entirely new images from random polygon inputs, translating images from external datasets into H\&E stained images, and applying GANs for style transfer, further expanding the versatility and application scope of these networks in diverse imaging contexts.

\begin{equation}
    \mathscr{L}_{GAN} = E_x[\log (\mathscr{D}(x)] + E_z[\log (1-\mathscr{G}(z)]
    \label{eq:gan}
\end{equation}

Alongside traditional augmentation techniques, \citet{60} also explore the use of four Deep Convolutional GANs (DCGANs)~\citep{dcgan} (one for each class) and a conditional GAN (cGAN)~\citep{cgan} to synthesize new images for classifying Gleason patterns. Additionally, SMOTE~\citep{smote} is employed to generate extra feature vectors that help train the fully-connected layers independently. \citet{136}  introduce an image translation approach using the CycleGAN~\citep{cyclegan}, which transforms images of normal colonic mucosa into less common polyp classes. They also implement a filtration step named the Path-rank Filter, involving a ResNet model to differentiate between real and synthetic images, selecting those with the most significant features based on output probability. The authors further explore translating natural images to histopathology images but find that intra-domain translation yields more realistic and higher-quality results. 

\begin{figure*}[t]
    \centering
    \includegraphics[width=\textwidth]{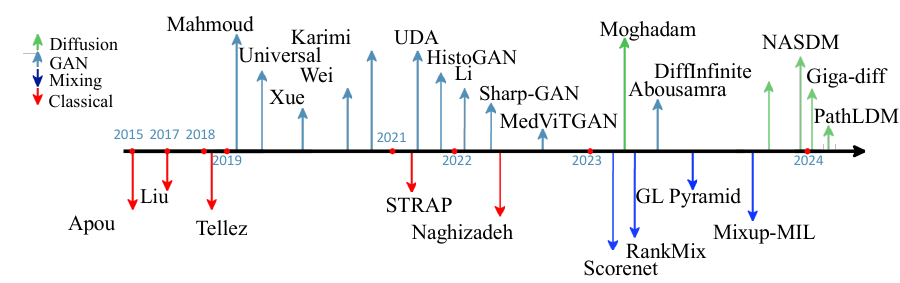}
    \caption{Chronological Trend Line of the Most Relevant Augmentation Techniques. The length of an arrow is proportional to its popularity per its category. GANs are the most popular choice for augmentation, with diffusion models gaining popularity recently.}
    \label{fig:aug_trend}
\end{figure*}
\citet{59} synthesize images for cancerous types for which no real images exist. Initially, quick, low-quality histopathology images are crafted by incorporating randomly generated polygons, colored with nuclear textures, into a tissue background. Subsequently, a GAN is employed to refine these synthesized images, while optimizing for the quality of the images. \citet{69} propose a centroid-based filtering process that selects informational and diverse images after using the cGAN. These images reside near their class's feature average. The discriminator of the cGAN incorporates a minibatch discrimination module~\citep{minidisc} to reduce similarity among generated images.

\citet{49} aim to generate images with perfect segmentation masks to improve the segmentation of highly-cluttered nuclei, and to develop a general learner robust against variation in the organ, staining, clinic, and disease state, while avoiding the collection of such highly diversified data. Image synthesis starts from random shapes and locations of polygons, which are conferred with contextual realism through a CycleGAN. Content and style of histopathology images are separated by \citet{51} via projection unto low-dimensional axes to reflect staining/equipment variation, and variation due to the hidden taxonomy of the disease. Consequently, cross-combining content and style axes generate new labeled histopathology images.

\citet{64} employ the CycleGAN or StarGAN~\citep{starGAN} to translate the stain of an existing labeled histopathology image into various other stains, thereby both diversifying and increasing the quantity of available data. Their method, named Unsupervised Domain Augmentation (UDA), retains the F-score performance with only 20\% of the training data. \citet{66} address ambiguity and image quality in their synthesized images with a selection mechanism based on the minimum expected predictive entropy and feature similarity with images in the same category. Minibatch discrimination complemented by a smoothed Fréchet Inception Distance (FID)~\citep{fid} is used to continuously assess the quality of the generated images.

\citet{47} develop Sharp-GAN, wherein conditioning occurs over distance maps derived from randomized polygons rather than traditional segmentation masks, thus producing images with clearly delineated nuclei. \citet{65} pair a generated image with a corresponding mask created by a segmentation head. The discriminator in the GAN is trained to differentiate synthetic from real pairs, penalizing both inadequate segmentations and subpar image synthesis. This process is conducted in a pyramidal fashion to ensure that features at various levels are accurately captured. 

MedViTGAN is transformer-based, and generates synthetic images in an end-to-end manner as proposed by \citet{48}. The reported synthetic images are diverse and of high-fidelity and their introduction to the training data of different learners was shown to improve their performance across two histopathological datasets. \citet{157} guide the image generation by spatial and topological analyses. Namely, the K-function~\citep{kfunc} and persistent homology~\citep{topo} descriptors are added as conditions to the GAN as well as in the loss function. This is in contrast to most previous methods relying on randomized polygons for histopathological objects. The incorporation of said descriptors in the loss also ensures that generated images match their original peers in holes, shapes, and density. 

\begin{table*}[]
\centering
\caption{Summary of Relevant Data Augmentation Methods in Histopathology, organized chronologically in descending order. (-) indicates unreported value. Numeric values are calculated with respect to the used baseline. Repeated datasets are in bold font.}
\label{tab:aug}
\begin{tabular}{>{\footnotesize}c>{\footnotesize}c>{\footnotesize}c>{\footnotesize}c>{\footnotesize}c>{\footnotesize}c>{\footnotesize}l}
\textbf{Type} & \textbf{Work} & \textbf{Task} & \textbf{Dataset} & \textbf{Annot. \%} & \textbf{Perf. \%} & \textbf{Quality} \\ \hline
\multirow{10}{*}{\rotatebox[origin=c]{45}{Diffusion}} & PathLDM - \citep{197} & CLS & \textbf{BRCA-M2C - Breast} & 0\% & \textcolor{red}{$\downarrow$}7.74\% (ACC) & 7.64 (FID) \\ \addlinespace 
 & Giga-diff - \citep{giga-diff} & - & - & - & - & - \\ \addlinespace 
 & Moghadam - \citep{morphology} & - & TCGA - Brain & - & - & 20.11 (FID) \\ \addlinespace 
 & \multirow{6}{*}{DiffInfinite - \citep{diff-inf}} & SEG & Private-2 & Zero-shot & \textcolor{green}{$\uparrow$}15.63\%   (D) & \multicolumn{1}{c}{-} \\
 &  & \multirow{5}{*}{CLS} & Private-1 & 100\% & \textcolor{green}{$\uparrow$}0.7\%   (ACC) & \multicolumn{1}{c}{26.7 (FID)} \\
 &  &  & Private-2 & \multirow{4}{*}{Zero-shot} & \textcolor{red}{$\downarrow$}0.2\%   (ACC) & \multicolumn{1}{c}{-} \\
 &  &  & Private-3 &  & \textcolor{green}{$\uparrow$}6.5\%   (ACC) & \multicolumn{1}{c}{-} \\
 &  &  & PCam - Breast &  & \textcolor{green}{$\uparrow$}2.07\%   (ACC) & \multicolumn{1}{c}{-} \\
 &  &  & NCT-CRC-HE - Colorectal &  & \textcolor{red}{$\downarrow$}1.25\%   (ACC) & \multicolumn{1}{c}{-} \\ \addlinespace 
 & NASDM - \citep{nasdm} & - & Lizard & - & - & \multicolumn{1}{c}{14.1 (FID)} \\ \hline
\multirow{6}{*}{\rotatebox[origin=c]{45}{Mixing}} & \multirow{2}{*}{MixUP-MIL - \citep{mixup}} & \multirow{2}{*}{CLS} & Thyroid - Frozen & \multirow{2}{*}{100\%} & \textcolor{green}{$\uparrow$}6.85\% (ACC) & - \\
 &  &  & Thyroid - Paraffin &  & \textcolor{green}{$\uparrow$}1.2\% (ACC) & - \\ \addlinespace 
 & GL Pyramid - \citep{163} & CLS & \textbf{BreaKHis - Breast} & 100\% & \begin{tabular}[c]{@{}c@{}}\textcolor{green}{$\uparrow$}5.54\% (ACC)\\ \textcolor{green}{$\uparrow$}2.28\% (D)\end{tabular} & - \\ \addlinespace 
 & Scorenet - \citep{131} & CLS & BRACS - Breast & 20\% & \textcolor{red}{$\downarrow$}6.38\% (F) & - \\ \addlinespace 
 & \multirow{2}{*}{RankMix - \citep{160}} & \multirow{2}{*}{CLS} & Camelyon16 - Breast & \multirow{2}{*}{100\%} & \begin{tabular}[c]{@{}c@{}}\textcolor{green}{$\uparrow$}2.6\% (ACC)\\ \textcolor{green}{$\uparrow$}0.02\% (AUC)\end{tabular} & \multirow{2}{*}{-} \\ \addlinespace 
 &  &  & TCGA - Lung &  & \begin{tabular}[c]{@{}c@{}}\textcolor{green}{$\uparrow$}0.51\% (ACC) \\ \textcolor{green}{$\uparrow$}0.15\% (AUC)\end{tabular} &  \\ \hline
\multirow{15}{*}{\rotatebox[origin=c]{45}{GAN}} & Abousamra - \citep{157} & CLS & \textbf{BRCA-M2C - Breast} & 100\% & \textcolor{green}{$\uparrow$}1.37\% (F) & - \\ \addlinespace 
 & Sharp-GAN - \citep{47} & SEG & \textbf{MICCAI15\&17} & 0\% & \textcolor{red}{$\downarrow$}8.06\% (J) & \begin{tabular}[c]{@{}l@{}}0.756 (S)\\ 0.793 (FS)\\ 0.147 (G)\\ 0.23 (E)\end{tabular} \\ \addlinespace 
 & \multirow{2}{*}{MedViTGAN - \citep{48}} & \multirow{2}{*}{CLS} & PCam - Breast & \multirow{2}{*}{100\%} & \begin{tabular}[c]{@{}c@{}}\textcolor{green}{$\uparrow$}6.58\% (ACC) \\ \textcolor{green}{$\uparrow$}4.26\% (AUC)\end{tabular} & 58.2 (FI) \\
 &  &  & \textbf{BreaKHis - Breast} &  & \begin{tabular}[c]{@{}c@{}}\textcolor{green}{$\uparrow$}3.44\% (ACC) \\ \textcolor{green}{$\uparrow$}5.04\% (AUC)\end{tabular} & - \\ \addlinespace 
 & \multirow{2}{*}{Li - \citep{65}} & \multirow{2}{*}{SEG} & GlaS & \multirow{2}{*}{20\%} & \textcolor{red}{$\downarrow$}6.02\% (D) & 0.0059 (FI) \\
 &  &  & Prostate &  & \textcolor{green}{$\uparrow$}5.1\% (J) & 0.013 (FI) \\ \addlinespace 
 & UDA - \citep{64} & SEG & Private (multi-stain) - Kidney & 20\% & \textcolor{red}{$\downarrow$}5.49\% (D) & - \\ \addlinespace 
 & HistoGAN - \citep{66} & CLS & Private - Cervical & 100\% & \begin{tabular}[c]{@{}c@{}}\textcolor{green}{$\uparrow$}8.89\% (ACC) \\ \textcolor{green}{$\uparrow$}5.38\% (AUC)\end{tabular} & - \\ \addlinespace 
 & Wei - \citep{136} & CLS & Private - Colorectal & 0\% & \textcolor{red}{$\downarrow$}12.83\% (AUC) & - \\ \addlinespace 
 & Mahmood - \citep{49} & SEG & Monu - Multi-organ & 100\% & \begin{tabular}[c]{@{}c@{}}\textcolor{green}{$\uparrow$}45.07\% (J) \\ \textcolor{green}{$\uparrow$}13.8\% (D)\end{tabular} & - \\ \addlinespace 
 & Karimi - \citep{60} & CLS & Private - Prostate & 100\% & \textcolor{green}{$\uparrow$}26.03\% (ACC) & - \\ \addlinespace 
 & Xue - \citep{69} & CLS & Private - Cervical & 100\% & \begin{tabular}[c]{@{}c@{}}\textcolor{green}{$\uparrow$}8.14\% (ACC) \\ \textcolor{green}{$\uparrow$}4.65\% (AUC)\end{tabular} & - \\ \addlinespace 
 & \multirow{3}{*}{Universal - \citep{59}} & \multirow{2}{*}{SEG} & \textbf{MICCAI17} & \multirow{3}{*}{0\%} & \textcolor{red}{$\downarrow$}1.31\% (D) & - \\
 &  &  & MICCAI18 &  & \textcolor{green}{$\uparrow$}0.1\% (D) & - \\
 &  & DET & Lymhpocyte Detection &  & \textcolor{green}{$\uparrow$}3.23\% (D) & - \\ \hline
\multirow{3}{*}{\rotatebox[origin=c]{45}{Classical}} & Naghizadeh - \citep{187} & - & Microscopical Images & N/A & N/A & N/A \\ \addlinespace 
 & \multirow{2}{*}{STRAP - \citep{62}} & \multirow{2}{*}{CLS} & CRC-DXT-TEST & \multirow{2}{*}{100\%} & \textcolor{green}{$\uparrow$}14.51\% (AUC) & - \\
 &  &  & Camelyon17-WILDS &  & \begin{tabular}[c]{@{}c@{}}\textcolor{green}{$\uparrow$}48.49\% (ACC) \\ \textcolor{green}{$\uparrow$}14.2\% (AUC)\end{tabular} & -
\end{tabular}
\end{table*}


\subsubsection{Mixing Techniques}
Several mixing techniques are proposed to synthesize new histopathology images. In general, the mixing process can be described as the linear combination 
\begin{equation}
    \lambda' = \alpha \cdot \lambda_i + (1-\alpha) \cdot \lambda_j
\end{equation} 
Herein, $\lambda$ represents the mixing element, which may be an image, a feature vector, or a label. The $i,j$ denote the mixed pair which generated the new augmented datum. Usually, $\alpha \stackrel{}{\sim} U(0,1)$. Mixing is illustrated for in Fig. \ref{fig:aug-mix}.

Mixup-MIL by ~\citet{mixup} consist of multi-linear mixing in addition to the vanilla linear mixing as in the equation above. Concretely, mixing is applied to random WSI patches of the same/different classes that are embedded through a feature extractor. The new label is obtained, similarly, as a linear combination of the original labels in case mixing occurs with different labels. The multi-linear combination regards $\alpha$ as vector-wise factor, and applies element-wise multiplication with the feature vector, thus introducing a higher degree of control for mixing. 

RankMix of \citet{160} precedes the mixing by a ranking process to determine the representativeness of extracted features. This ranking can be accomplished through various methods, such as utilizing a MLP for patch-level pseudo labeling or employing attention-based MIL. In contrast, \citet{131} introduce ScoreMix, which operates directly in the image domain. This method utilizes a ViT to identify ROIs that significantly contribute to the representation of a WSI, denoted as $X^{(1)}$. Selected ROIs, $ROI_i$, are transposed from $X^{(1)}$ to an uninformative patch, $X^{(2)}$, creating a new and unique patch. Additionally, the labels from the original WSIs are blended according to mixing equation above using weights $\lambda_i = Y_i$ and $\lambda_j = Y_j$, thus forming a composite label for the newly synthesized image.

\citet{163} also explore a mixing strategy, but their approach combines halves of different images. This method incorporates Gaussian-Laplacian (GL) pyramid blending to smooth out color and staining discrepancies between the combined image sections, thus enhancing the visual consistency of the resultant WSI. 



\subsubsection{Diffusion Models}
Probabilistic Diffusion Models (PDMs)~\citep{pdm1,pdm2} have been introduced to overcome some of the limitations associated with GANs, such as mode collapse and training instability. PDMs generally operate through a two-step process: the forward diffusion phase, where noise is incrementally added to an image, and the reverse diffusion phase, which is trained to methodically remove this noise, effectively reconstructing the original image from its noised state, and potentially conditioned on various types of conditions. This is illustrated for in Fig. \ref{fig:aug-diff}, where the condition is a segmentation mask. \citet{multimodal} delve into the efficacy of PDMs compared to GANs in generating histopathology images, presenting results that suggest PDMs may offer superior performance. Akin to cGANs, diffusion models often incorporate conditions into the synthesis process to enhance the quality, diversity, and overall fidelity of the generated images. This conditioning helps tailor the output to specific requirements and improves the utility of the synthesized images for downstream applications. Moreover, contributions in this line of work are able to synthesize high-resolution images, potentially up to the scale of WSIs, in contrasts to the limited synthesis resolution observed in most GAN-based approaches.

\begin{figure*}[t]
    \centering
    \includegraphics[width=6 in]{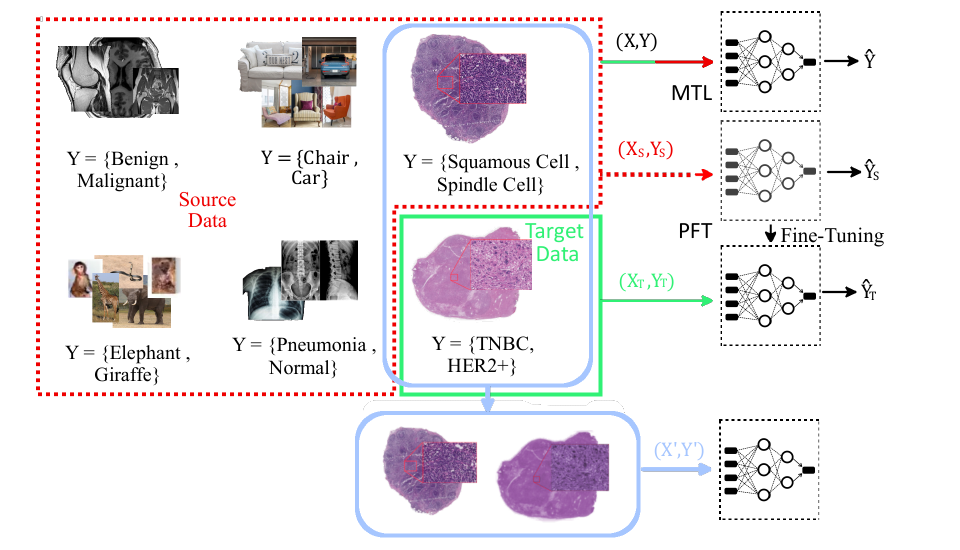}
    \caption{Illustration TL paradigms. In MTL, all of $(X,Y)$ dataset, including source domain (in red) and target domain images (in green) are used to train the learner simultaneously. In pretraining plus fine-tuning, initial knowledge is built using the source domain data (red dashed lines ($X_s,Y_s$) and is then fine-tuned on the target domain data (green line ($X_T,Y_T$). In blue: an example of domain adaptation: involved domains are projected into a color space which minimizes their shift. Images in the source domain(s) are assumed plenty, and their labels easier to obtain.}
    \label{fig:fine}
\end{figure*}

\citet{morphology} introduce a PDM specifically conditioned on the genotype of the image. The used color normalization nullifies the color effect and focuses the synthesis on the morphological features. Genotypes are incorporated into the model via an embedding layer, allowing the network to integrate genetic information directly into the image generation.

In a related approach, \citet{197} utilize a Variational Autoencoder (VAE) for image generation, conditioning the process on textual descriptions of the images. These descriptions are processed through a CLIP~\citep{clip} model to encode the text. Their experiments reveal that using non-relevant textual information, or omitting textual context altogether, significantly degrades the quality of the generated images.

\citet{giga-diff} propose a novel coarse-to-fine diffusion model designed to synthesize Giga-pixel scale histopathology images. This framework establishes the coarse structure of the image in the initial stages, with progressively finer details added to ensure that both global and local features are accurately represented. The reverse diffusion process in this model is uniquely conditioned on the spatial resolution of the image, which is derived from the metadata of the WSI, optimizing the synthesis for high-resolution outputs.

The DiffInifine model proposed by \citet{diff-inf} and the Nuclei-aware semantic tissue generation diffusion model (NASDM) by \citet{nasdm} both have the advantage of conditioning the reverse-diffusion step on a segmentation mask, thus creating synthetic mask-image pairs, and greatly increasing the utility of produced images in the segmentation/detection tasks. Both models also make use of the semi-supervised classifier-free guidance \citep{free-guidance}, which occasionally switches to non-conditional training; therefore, making use of unlabeled data to improve the diversity of synthetic images.

The NASDM model separates the segmentation mask per nucleus class, and adds another mask consisting of nuclei edges, and injects said information into the decoder of the denoising network. The DiffInfinite model is additionally capable of producing high-resolution images. It extracts small patches from the latent space of training WSIs, upon which multiple realizations of a diffusion model are established. Firstly, high-level, low-frequency aspects of the image (such as a sketch of the cellular arrangement) are learned conditioned on simple textual prompts. These masks can be upsampled linearly to high resolutions. Then, small crops are randomly extracted from large images, and correspondingly from the generated mask to learn the synthesis of low-level, high-frequency images conditioned on the segmentation mask. The authors detail the combination process of the various crops to obtain high-resolution images with their corresponding masks. Generating image-mask pairs.


\subsubsection{Summary}
We provide a comparison of approaches in the literature in Table.~\ref{tab:aug}, in addition to a chronological summary of these methods in Fig.~\ref{fig:aug_trend}. In the following points, we summarize key points about data augmentation in the literature of histopathology.

\begin{itemize}
    \item Data augmentation methods are designed to increase the training data pool by generating new, ideally labeled, instances.
    \item Traditional augmentation techniques are increasingly complemented by or replaced with GANs, which are favored in recent studies over classical approaches due to their high fidelity and suitability for histopathology images.
    \item In the domain of histopathology, CycleGANs and conditional GANs are commonly employed across various studies.
    \item Recent advancements in the field also incorporate diffusion models due to their ability to solve some shortcomings faced by GANs, and integrate textual data into the augmentation processes.  
    \item Synthesizing images conditioned on segmentation masks increases the utility of generated images by allowing segmentation models to train on them. Accordingly, diffusion models that follow this principle can assess their models on downstream segmentation tasks.
    \item A promising area for exploration is the use of textual information for image augmentation in histopathology. Currently, this approach is underexplored, despite its potential to enhance augmentation techniques significantly.
\end{itemize}


\section{Transfer Learning}
\label{section:TL}

Transfer learning (TL) \citep{transfersurvey,transfersurvey2} techniques address the problem of data scarcity by outsourcing knowledge, skill and representation learning to foreign data. Let $(X^{WSI},Y^{WSI})$ represent the labeled data in the targeted histopathology domain. With TL, we seek to expand the training data $X$, such that $(X,Y) = (X^{WSI},Y^{WSI}) \cup (X^{WSI,k},Y^{WSI,k}) \cup (X^{non-WSI},Y^{non-WSI})  \supset (X^{WSI},Y^{WSI})$. As such, the ramifications of annotation scarcity are mitigated, specifically through enhancing the generality of the learned features by training on data sets that are easier collect and/or annotate. The knowledge learned from the source task(s) is used to improve the performance of the model on the target task, despite the small scale of the target task labels, or lack thereof. We categorize TL methods into fine-tuning, multitask learning, and domain adaptation techniques. Our division of TL methods is principally consistent with previous ones \citep{transfersurvey}, yet we amend some details to tailor it to the reviewed works in pathology.

\begin{figure*}[t]
    \centering
    \includegraphics[]{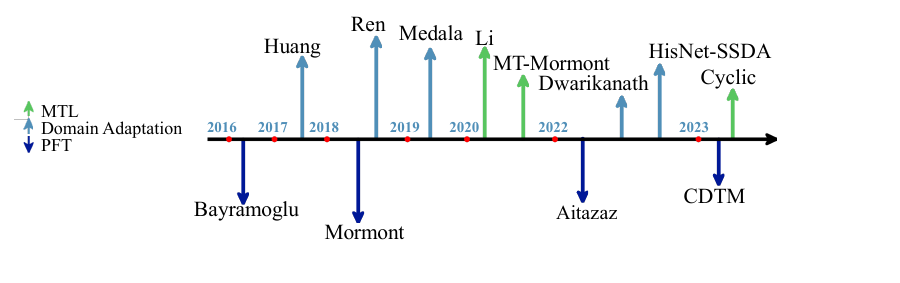}
    \caption{Chronological Trend Line of the Most Relevant Transfer Learning Techniques. The length of an arrow is proportional to its popularity per its category.}
    \label{fig:transfer_trend}
\end{figure*}

\subsection{pre-training and fine-tuning}
pre-training plus fine-tuning (PFT) methods form a sub-category of TL methods, whereby a model is firstly pre-trained on large labeled data, such as natural images, or datasets combining several types of cancer different from the targeted cancer type. The fine-tuning procedure then modifies some of the model's weights, usually in the final layers, by training on the small-scale targeted data. Pre-training teaches the learner meaningful features, which the targeted cancer type may share. Even with natural images, it has been demonstrated that models can learn useful features~\citep{imagenet2,imagenet1,80,73}.




The work of \citet{98}, \citet{181}, \citet{191}, and \citet{182} exemplify the earlier trends in PFT approaches. Principally, CNN architectures (AlexNet, GenderNet, GoogLeNet, VGG-16, and Inception) are trained on natural and facial images, and are adapted to learn histopathology tasks by fine-tuning and adjusting learning rates differently (as in~\citep{98}) for layers from the source network and new layers in the target network. The results demonstrate that PFT generally outperforms full training on the small-scale medical data. Later works~\citep{87} propose the usage of the ViT for feature extraction and demonstrate its superior performance to the CNN.

In~\citep{97}, a deeper analysis of the representation transferability from the natural domain to the medical domain is conducted. The authors define the information gain of a layer in a pre-trained CNN by the gap between its classification accuracy and that of a random-weight layer. The study concludes that middle-layer representations lead to the highest diagnosis rates, while the transferred general knowledge mainly resides in early layers. Other works affirm that employing a histpathology source domain for pre-training followed by fine-tuning the model on a histopathological data from the target domain results in superior performance as compared to transferring from natural images~\citep{79, 80}. Nonetheless, a recent work by \citet{190} makes use of LAION-2B, consisting of a billion-level natural image-text pairs~\citep{laion5b}, following the PFT paradigm and adding an extra learnable module dubbed CDTM (cross-domain transfer model). CDTM filters the output feature of the pre-trained layers, thus preventing gradient explosion/vanishing, and assimilating the model to the target domain. They also propose to gradually unfreeze the learner where the pre-trained layers are firstly frozen in the first stage while fine-tuning the CDTM and linear head, and are then unfrozen where the entire model is fine-tuned.

\subsection{Multitask learning}
In contrast to PFT, Multi-task learning (MTL) methods train a model to perform multiple related tasks simultaneously. The expected advantage is an implicit regularization that improves convergence and reduces overfitting. In concrete terms, when multiple tasks are learned together the training will be less prone to overfitting either task; instead, the learner may learn general features suitable to all tasks at once, increasing the possibility of generalizing said features to unseen data. 
 
\citet{75} propose a MTL framework, where they unify multiple histopathological datasets into a training pool of around 900 thousand images constituting 22 different classification tasks. A shared network is used for feature extraction and is attached to several fully connected networks; one for each task. The fully connected networks have few parameters to concentrate the representation learning on the shared network. \citet{mtl1} train a learner to simultaneously optimize a classification loss and a verification loss. The objective of the latter is to decrease intra-class distances and increase inter-class distances between a pair of images. \citet{145} train a classifier concurrently with a nuceli segmentor in a cyclic learning paradigm. Starting with image classification, a layer-wise relevance propagation (LPR)~\citep{LPR} module localizes the prediction into an incomplete pseudo mask, which in turn is trained in a teacher-student framework to learn from partially labeled data. Cyclically, the weights between the two learners are regarded as initialization of the other. 

In~\citep{132}, a classifier is trained to predict the Whole-Genome Doubling (WGD) phenotype from histopathology images of several types of cancer. To that end, the Model-Agnostic Meta-Learning (MAML) trains a learner to extract representations from a few images representing several cancer types. The meta-training data consists of Bladder Urothelial, Breast, Colon, and Head and Neck carcinomas, in addition to Lung adenocarcinoma, Lung squamous cell carcinoma, Rectum adenocarcinoma and Stomach adenocarcinoma. A meta-validation set is designated to tune the hyperparameters of the model and includes Esophageal, and Liver hepatocellular carcinomas. The meta-testing set includes Adrenocortical carcinoma, Cholangiocarcinoma, Kidney Chromophobe, Kidney renal clear cell carcinoma, Ovarian serous cystadenocarcinoma, Uterine Carcinosarcoma, and Uterine Corpus Endometrial Carcinoma. By employing a meta-learning scheme, the learner ability to overfit a specific cancer type is curbed, despite only relying on a few examples during training.

\subsection{Domain Adaptation}
Works such as~\citep{79,75,80,145} and others importantly highlight that the closer the source domain is to the target domain (or the multiple tasks learned simultaneously), the more consistent, and hence the more useful the learned representations are. Consequently, many approaches apply TL while lending a special care to the issue of domain adaptation. Domain adaptation methods focus on addressing the differences between the source and target domains, which pertain to 1) the statistical distributions of the domains, or 2) their label set. On one hand, domain adaptation confers learners with the ability to borrow learned representations from foreign domains, while on the other, it enhances the applicability of representations learned from the training data (source domain) to external test sets (target domains), thereby obviating the need for larger data cohorts. This makes domain adaptation highly relevant in the context of data scarcity. Herein, semi/un-supervised domain adaptation refers to the annotation content in the \textit{target} domain. 

\begin{table*}[]
\centering
\caption{Summary of TL methods per type, organized chronologically in descending order. Annotation and Performance percentages are calculated w.r.t to the mentioned baseline in each work. Common datasets are in boldface.}
\label{tab:transfer}
\begin{tabular}{>{\footnotesize}c>{\footnotesize}l>{\footnotesize}c>{\footnotesize}c>{\footnotesize}c>{\footnotesize}l}
\textbf{Type} & \multicolumn{1}{c}{\textbf{Work - Year}} & \textbf{Task} & \textbf{Dataset - Organ} & \multicolumn{1}{l}{\textbf{Annot. \%}} & \multicolumn{1}{c}{\textbf{Perf. \%}} \\ \hline
\multirow{8}{*}{\rotatebox[origin=c]{45}{Domain Adaptation}} & Dwarikanath - \citep{188} & CLS & Camelyon17 - Breast & N/A & \textcolor{red}{$\downarrow$}0.72\% (AUC) \\ \addlinespace 
 & \multirow{2}{*}{HisNet-SSDA - \citep{82}} & \multirow{2}{*}{CLS} & Zhejiang University & \multirow{2}{*}{16.7\%} & \textcolor{red}{$\downarrow$}2.29\% (ACC) \\
 &  &  & Digestive-Sys, Path. Det.and Seg. Challenge &  & \textcolor{red}{$\downarrow$}1.78\% (ACC) \\ \addlinespace 
 & Medala - \citep{193} & CLS & B-CBL-HE & 20\% & \textcolor{green}{$\uparrow$}9.76\% (ACC) \\ \addlinespace 
 & \multirow{2}{*}{Ren - \citep{194}} & \multirow{2}{*}{CLS} & Private - Prostate & \multirow{2}{*}{100\%} & \textcolor{green}{$\uparrow$}33.21\% (ACC) \\
 &  &  & TCGA - Prostate &  & \textcolor{green}{$\uparrow$}41.99\% (ACC) \\ \addlinespace 
 & \multirow{2}{*}{Huang - \citep{185}} & \multirow{2}{*}{CLS} & VGH & \multirow{2}{*}{100\%} & \textcolor{red}{$\downarrow$}0\% (ACC) \\
 &  &  & NKI &  & \textcolor{green}{$\uparrow$}2.06\% (ACC) \\ \hline
\multirow{10}{*}{\rotatebox[origin=c]{45}{MTL}} & \multirow{3}{*}{Cyclic - \citep{145}} & \multirow{3}{*}{SEG} & MoNu & \multirow{3}{*}{CL} & \begin{tabular}[c]{@{}l@{}}\textcolor{red}{$\downarrow$}0.9\% (D)\\ \textcolor{red}{$\downarrow$}1.29\% (AJI)\end{tabular} \\ \addlinespace 
 &  &  & CCRCC &  & \begin{tabular}[c]{@{}l@{}}\textcolor{red}{$\downarrow$}0.3\% (D)\\ \textcolor{red}{$\downarrow$}0.51\% (AJI)\end{tabular} \\ \addlinespace 
 &  &  & CoNSeP &  & \begin{tabular}[c]{@{}l@{}}\textcolor{red}{$\downarrow$}4.6\% (D)\\ \textcolor{red}{$\downarrow$}8.91\% (AJI)\end{tabular} \\ \addlinespace 
 & \multirow{2}{*}{Li - \citep{mtl1}} & \multirow{2}{*}{CLS} & \textbf{BreaKHis - Breast} & \multirow{2}{*}{100\%} & \textcolor{green}{$\uparrow$}8.31\% (ACC) \\
 &  &  & Lymphoma subtyping &  & \textcolor{green}{$\uparrow$}7.11\% (ACC) \\ \addlinespace 
 & \multirow{5}{*}{MT-Mormont - \citep{75}} & \multirow{5}{*}{CLS} & BoneMarrow & \multirow{5}{*}{100\%} & \textcolor{green}{$\uparrow$}21.42\% (ACC) \\
 &  &  & Breast1 &  & \textcolor{green}{$\uparrow$}2.59\% (AUC) \\
 &  &  & Breast2 &  & \textcolor{green}{$\uparrow$}5.46\% (AUC) \\
 &  &  & \textbf{Kidney} &  & \textcolor{green}{$\uparrow$}1.01\% (AUC) \\
 &  &  & Lung &  & \textcolor{green}{$\uparrow$}6.32\% (ACC) \\ \hline
\multirow{7}{*}{\rotatebox[origin=c]{45}{PFT}} & \multirow{2}{*}{CDTM - \citep{190}} & \multirow{2}{*}{CLS} & \textbf{BreaKHis - Breast} & \multirow{2}{*}{100\%} & \begin{tabular}[c]{@{}l@{}}\textcolor{green}{$\uparrow$}6.96\% (ACC)\\ \textcolor{green}{$\uparrow$}10.1\% (F)\end{tabular} \\ \addlinespace 
 &  &  & HCRF - Gastric &  & \begin{tabular}[c]{@{}l@{}}\textcolor{green}{$\uparrow$}1.34\% (ACC)\\ \textcolor{green}{$\uparrow$}1.34\% (F)\end{tabular} \\ \addlinespace 
 & Aitazaz - \citep{87} & CLS & LC25000 - Colon/Lung & 100\% & \textcolor{green}{$\uparrow$}10.99\% (ACC) \\ \addlinespace 
 & \multirow{3}{*}{Mormont - \citep{182}} & \multirow{3}{*}{CLS} & Breast & \multirow{3}{*}{100\%} & \textcolor{green}{$\uparrow$}5.65\% (AUC) \\
 &  &  & Lung &  & \textcolor{green}{$\uparrow$}10.04\% (AUC) \\
 &  &  & \textbf{Kidney} &  & \textcolor{green}{$\uparrow$}4.46\% (AUC) \\ \addlinespace 
 & Bayramoglu - \citep{98} & CLS & HistoPhenoptypes - Multi-organ & 100\% & \textcolor{green}{$\uparrow$}0.87\% (ACC)
\end{tabular}
\end{table*}

In~\citep{185}, authors project both source and target data into a common subspace using unsupervised Principal Component Analysis (PCA), which minimizes the domain shift between the two domains. Notably, the method utilizes a closed-form analytical solution in place of the traditional iterative back-propagation. \citet{194} reduce the shift between the source and target domains in a Gleason grading task. A target network mimics the features extracted by a source network, where the latter is trained on the labeled, source data. A Siamese branch in the target network encourages patches from the same WSI to have the same Gleason. The authors later~\citep{192} additionally investigated another means of unsupervised domain adaptation via color normalization between the domains, but it was reported that the adversarial approach is superior. 

\citet{193} reduce the shift between the source data constituting complete, well-labeled colorectal cancer tissue samples and the target data consisting of a few images of colon, breast and lung tissue patches. The Siamese NN \citep{snn} is trained on the source data using the Triplet Loss~\citep{triplet}, which is then used as a feature extractor connected to a shallow classifier and trained on few samples in the target domain. \citet{107} apply image translation from the source flourecense to the target histoapthology images using a CycleGAN improved by a nuclei inpainting step to remove unwanted auxiliary nuclei in the generated images. Moreover, the authors establish a semantic-level adaptation between the synthesized and real images, by learning domain-invariant features at the semantic level. Furthermore, a task re-weighting protocol is used to ensure that the learned features are domain-invariant, thus promoting domain adaptation. This is done by decreasing the importance of domain-specific and source-biased features, and increasing the importance of hard-to-differentiate features.


\citet{82} apply semi-supervised domain adaptation between a label-rich source domain and a target domain which is partially annotated. Adaptation is based on a proposed weight function consisting of the cross-entropy loss over labeled samples in both domains, an extension to it called the unlabeled conditional cross-entropy applied to unlabeled samples, and a regularization term named the maximum mean discrepancy (MMD) to model the difference between the domains in high-dimensional feature space. \citet{188} propose to disentangle learned features into domain-invariant (structural) and domain-variant (textural or stylistic) features. Here, the structure from the source domain is combined with the content from the target domain through a decoder-encoder chain, thus creating a shared representation among both domains.



\subsection{Summary}
Table.\ref{tab:transfer} compares different TL methods quantitatively, and we also provide a chronological summary of the methods in Fig.~\ref{fig:transfer_trend}. In the following, we provide the main key-points related to TL methods in histopathology.

\begin{itemize}
    \item Transfer learning, in general, refers to methods that aim to improve the performance of a learner by training it on external tasks/domains. A source domain is a domain that is easily accessible, and can help the model learn general features that can be transferred to the target domain.
    \item Transferring the knowledge from the source domain to the target domain often requires some conditioning, commonly referred to by domain adaptation.
    \item Transfer learning may take place sequentially (e.g. PFT methods) or simultaneously (e.g. MTL).
    \item It is mostly observed that a network pre-trained on natural images and/or histopathology images from various cancer types has a better training profile than a network trained from scratch on the target domain only. This is realized in the convergence and the overall error rate.
\end{itemize}

\section{Discussion, challenges, and Future Trends}
Principally, this review aims to devise the strategies to address the challenge of data scarcity and promote the clinical adoption of deep learning (DL) tools in digital pathology. We discussed data scarcity, one of the main aspects that hinders the development of DL tools in histopathology, and, as a result, said acceptance. Various other challenges are faced by researchers in the field working towards the clinical adoption of DL tools. These involve addressing the data scarcity factors, exploring alternative directions to the problem of data scarcity, decision interpretability, reproducability, and interoperability and standardization.

\subsection{Addressing Data Scarcity Factors}
In \ref{sec:motiv}, we fleshed out some of the factors causing a downward pressure on the compilation and sharing of pathology data, including the increased burden over clinicians, and the ethical aspects concerning the usage of sensitive medical data by researchers. Addressing these factors may have a cumulative benefit over the development of strong DL models. 

Firstly, crowd-sourcing is an imperative approach to the annotation of large datasets, as demonstrated across a wide range of applications in the natural scene. This remains challenging in the medical domain, as the latter requires trained personnel, in contrast to the natural scene where such a requirement is relaxed. Nonetheless, several analyses report promising results in this direction \citep{crowd-medical, crowd-survey}, even in the pathology domain \citep{crowd-pathology}.

Secondly, works such as \citep{ethics} argue that many of the ethical concerns surrounding the compilation and dissemination of WSIs data can be reconciled, such as the repurposing of tissue use for scanning and research, and the collection, handling, and storing of large medical records. Moreover, studies report promising areas of development in collaborative model training across multiple data centers, such as Federated Learning and Differential Privacy \citep{federated2,federated1,federated3}, or the decentralized blockchain-based Swarm Learning \citep{swarm}, which have been spurring much research due to their elevated privacy standards, and beneficial effect on performance. We encourage and expect that future directions would involve such tendencies.

Herein, we also encourage the pursuit of research that promotes the de-identification of patients, thus improving the privacy of involved subject studies, and diminishing the ethical concerns surrounding the art. Additionally, we encourage initiatives that raise awareness to the potential benefits of WSI research among patients, and other initiatives and acts that reassure the public about their privacy concerns.

\subsection{Efficient, Real-time DL-based CPath}
Large DL models do not only demand large datasets, which is the primary concern of this review, but also require expensive hardware, and extensive periods of time to train and use in inference mode. These factors may represent additional hurdles before the clinical acceptance of DL tools on a large scale. In this regard, quantization \citep{quantization}, pruning \citep{pruning}, and knowledge distillation \citep{distil} methods are among the leading research areas that address this issue \citep{efficient}.

\subsection{Decision Interpretability}
The problem of interpretable and explainable decision making in DL models is not unique to histopathology; yet, it is amplified. An explainable pipeline is essential to increase confidence in its decisions, facilitates further discovery, and improves integration within existing clinical practice. Within the context discussed in this review, it is noted that some methods are more interpretable than others. 

Decision-fusion MIL, segmentation, and detection tasks  are better explained than WSI classification within the feature-fusion setting. As a remedy, it is customary for most methods in the literature belonging to the latter to address this shortcoming by integrating CAM-based or attention-based localization. Similarly, data reduction methods promote explainability by delineating the most representative and descriminative regions

\subsection{Reproducibility}
Another important obstacle that may hinder the clinical adoption of DL tools is the issue of questioned reproducibility observed in a broad spectrum of the literature body. This issue may arise due to unshared details about training or pre-processing stpng, the employment of inconsistent testing protocols among researchers, or the usage of private datasets during training/testing, which is a recurring phenomenon that has been observed in this review as can be seen in benchmarking Tables \ref{tab:bench-mil},\ref{tab:semi},\ref{tab:aug}, and \ref{tab:transfer}.

Much of these factors may be addressed in cooperation by researchers, editors and reviewers of publication venues. Several recommendations are made including the deployment of large, realistic benchmarking datasets, the standardization of testing protocols, including performance metrics, and fixed training/testing splits, and the enforcement of detailed documentation of published approaches \citep{challenges}. To this end, we additionally propose the incentivisation of data sharing among researchers by addressing potential privacy and ethics concerns, field-wide recognition, and even monetary incentives.  

\subsection{Generalization Ability}
CPath applications cover a broad spectrum of use cases, yet many existing approaches are constrained by their reliance on specific domains or exhibit high effectiveness only within certain datasets. This limitation hinders the clinical adoption of these methods, as they need to accommodate diverse testing centers, equipment, and variations introduced by human factors. Overcoming these limitations is essential for advancing the art. This underscores the need for developing techniques that not only generalize well across diverse datasets but also exhibit zero-shot adaptability to various downstream tasks such as the many examples observed in this review.

\subsection{Future Trends}
Recent advancements in DL techniques addressing data scarcity emphasize localization support. This focus benefits the development of new and emerging techniques and aligns with the growing attention to interpretability in the medical practice. Incidentally, diffusion models have shown significant potential in data synthesis, particularly addressing the various challenges faced by GANs. However, we believe that research on synthesizing images with pixel-level annotation remains underexplored. There is also growing interest in multi-modal approaches, where language or genomic data is combined with histopathological images \citep{gen-path3,gen-path,gen-path2}. This direction is seen across the different methods discussed in this review, including semi-supervised learning, and data augmentation. Moreover, foundation models have recently demonstrated promising improvements, and is expected to drive further research. These large models, trained on extensive datasets, exhibit excellent generalization capabilities and hold significant potential for data-scarce applications, particularly in zero-shot and few-shot learning scenarios. 

Furthermore, we believe that the area of data reduction offers significant potential for further research and encourage scholars to explore this topic. Finally, the challenges discussed in this review, as well as those highlighted in other published works, are crucial for the clinical acceptance of the field and should be prioritized by researchers.





\section{Conclusion}
This review paper provides a comprehensive survey of state-of-the- DL techniques applied to computational histopathology, with a focus on addressing the critical challenge of data scarcity, a major barrier to the clinical adoption of DL tools. We present a logical and detailed taxonomy to facilitate the understanding, comparison, and contrast of various contributions made by researchers over the past decade, highlighting key themes and suggesting potential avenues for future development.
The collective results reported by different methods indicate encouraging progress toward the clinical integration of DL in histopathology. Accurate outcomes have been achieved in crucial tasks such as classification, detection, and segmentation, which mirror the diagnostic and prognostic practices regularly employed by pathologists. The field is rich and diverse, incorporating various techniques that optimize the use of noisy, scarce, coarse, external, synthetic, and alternative data sources. 

Recent contributions have demonstrated the advantages of integrating language and image representations. However, certain promising techniques, such as data reduction methods, remain understudied. Additionally, many contributions emphasize localization and interpretability, further supporting the clinical adoption of DL tools. Looking forward, we anticipate continued advancements in addressing data scarcity, as evidenced by the exponential growth in proposed methods over the past decade.

\section*{Declaration of Competing Interest}
The authors confirm that there are no competing financial, personal, or other types of interests that may influence the work reported in this paper, its conduction, preparation, or submission.

\section*{Acknowledgements}
This work is supported by research grants from the Advanced Technology Research Center Program (ASPIRE), Ref: AARE20-279, and the Terry Fox Foundation, Ref:I1037.

\bibliographystyle{elsarticle-harv}
\bibliography{survey_bibl}

\end{document}